# ASCEND: Automated Shockproof Cost-Efficient Network Design

Taha ValizadehAslani, Vaibhav Phatarpekar, Cameron Brackmann, Bala Ramachandran

***Abstract*—Designing resilient and cost-effective IP network topologies over an existing physical infrastructure is a challenging combinatorial optimization problem, particularly at large scale and under extensive failure modeling assumptions. Traditional deterministic approaches struggle to jointly address resiliency against shock (large customer-impact outages) and implementation cost objectives when the number of candidate circuits grows quadratically with the number of sites to be interconnected. In this paper, we propose Automated Shockproof Cost-Efficient Network Design (ASCEND), a systematic design pipeline for IP core network optimization based on a two-level optimization framework. At the outer level, ASCEND uses Multi-Objective Particle Swarm Optimization (MOPSO) to select core sites and core circuits. Given a core design, the inner-level integral Multi-Commodity Flow (MCF) formulation selects edge circuits and maps both core and edge circuits onto optical (fiber/transport) paths while prioritizing resource disjointness and latency. Network designs are evaluated using two competing objectives: a performance cost capturing resiliency under probabilistic failure scenarios and an implementation cost. The performance cost incorporates an exponentially weighted vulnerability metric that explicitly penalizes infeasible configurations. Experimental results on multiple real-world networks demonstrate that ASCEND consistently identifies designs that outperform a manually engineered baseline in both resiliency and implementation cost.**

## I. Introduction

As computer networks grow in size and complexity, flat non-hierarchical architectures can become increasingly difficult to scale and manage because routing state, control-plane interactions, and operational complexity grow with network size [1], [2], [3], [4], [5], [6]. Hierarchical network design addresses these challenges through a modular, layered approach that improves performance, enhances resiliency, and simplifies management [3], [4], [6]. Separating the network into core and edge segments helps networks operate more predictably, allowing for better fault isolation and easier scalability [4]. In this design, the metro core segment, consisting of high-speed network devices, is responsible for the transmission of data between the components of the edge/access segment (which, in turn, provides access for the users), and the service provider backbone network [4]. However, engineering an optimal IP (Layer-3) core topology over an existing fiber-based physical infrastructure (Layer-1) (see Fig. 1) remains a difficult combinatorial optimization problem: the number of candidate inter-core circuits grows quadratically with the number of candidate core sites, while deployment constraints (latency limits, capacity limits, device limits, and cost) sharply restrict feasible designs.

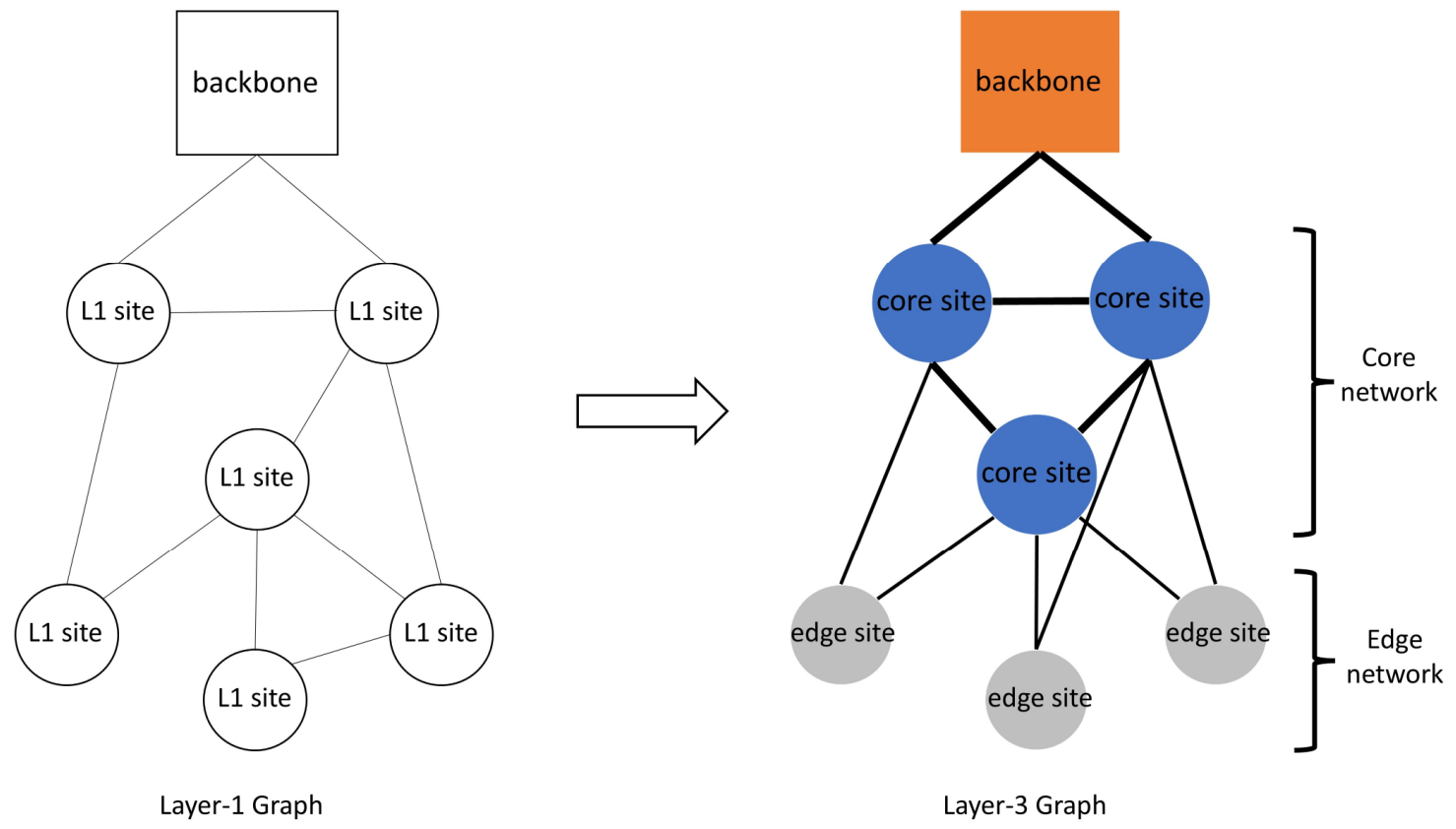

Fig. 1. Left: Layer-1 infrastructure, consisting of sites, backbone connectivity and fiber links. Right: Layer-3 (IP) network with a hierarchical design where some sites are core and some sites are edge. Each Layer-3 circuit consists of one or more Layer-1 links.

We use the term “shock” to denote relatively rare but high-impact outage outcomes, *i.e.*, failure scenarios that disconnect an unusually large number of customers. A major complication is that failures in real networks are frequently correlated rather than independent. Seemingly diverse logical circuits may share common physical risk factors such as conduits or geographically co-located fiber segments, so a single incident can disable multiple links simultaneously, which can cause a major shock to the

Comcast, Technology & Product NGAN, Philadelphia, PA, USA.

network. These correlated risk domains are commonly captured using the Shared Risk Link Group (SRLG) abstraction, and availability analyses show that network robustness can change substantially when SRLG failures are considered [7]. Importantly, resiliency cannot be assessed purely at the logical topology level: failures occur on fiber (Layer-1), but customer impact is realized at higher layers through the mapping of logical circuits onto physical paths. Consequently, effective core design requires physically grounded embeddings that account for shared capacity and explicitly seek diversity across links, nodes, and SRLGs, rather than relying merely on shortest-path routing that can inadvertently concentrate risk.

These observations motivate an automated design approach that treats IP topology synthesis and fiber/transport path (Layer-1) realizability as a coupled problem. Existing work addresses parts of this problem, either routing a given logical topology in a survivable manner over the physical layer or designing a topology without regard to its physical realization, but not the joint synthesis of an IP core and its fiber-level embedding under correlated failures (Section II).

In this paper, we address the following question: given an infrastructure of sites, fiber links, and SRLGs, how can we build an optimal IP hierarchical network design (core and edge connectivity) that jointly optimizes resiliency and implementation cost under various failure assumptions? To solve this problem, we propose ASCEND (Automated Shockproof Cost-Efficient Network Design), a two-level optimization framework for resilient and cost-efficient IP network design. At the outer level, ASCEND explores candidate selections of core sites and circuits using multi-objective particle swarm optimization (MOPSO) [8], [9], [10]. At the inner level, given a candidate core design, an integral Multi-Commodity Flow (MCF) formulation maps each core circuit to a feasible fiber-path realization, selects the edge circuits, and maps them onto fiber paths. The formulation prioritizes physical diversity (link/node/SRLG disjointness) while minimizing end-to-end delay under capacity constraints. Each design is evaluated using two competing objectives: a performance cost that captures resiliency under probabilistic failure scenarios (including SRLG correlated failures) and an implementation cost capturing deployment expense (see Fig. 2).

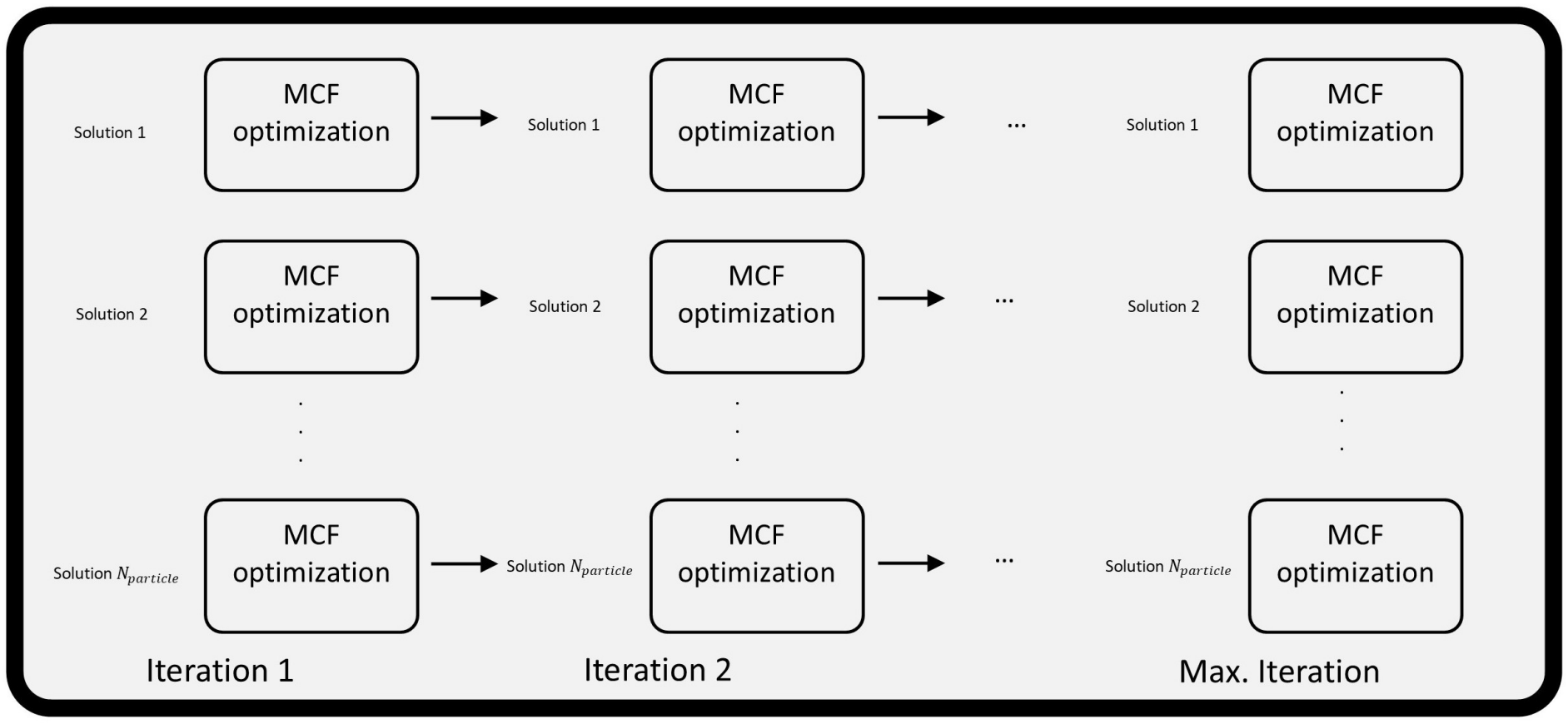

Fig. 2. Two-level optimization of ASCEND. Outer level: MOPSO candidate generation and update. Inner level: MCF embedding for each candidate.

This paper makes the following key contributions:

- **Problem formulation**: We formalize joint IP core site and core circuit selection over an existing fiber infrastructure under SRLG-aware correlated failure modeling.
- **Method**: We introduce ASCEND: A nested optimization pipeline that couples MOPSO-based topology synthesis with an integral MCF formulation to ensure physically realizable, diversity-aware designs.
- **Metric**: We propose a probabilistic vulnerability metric for analyzing network failure scenarios with an exponential outage penalty and explicit infeasibility penalties to reflect operational risk.
- **Evaluation**: We validate ASCEND on real-world scenarios and demonstrate improved Pareto tradeoffs between resiliency and implementation cost relative to an expert baseline. ASCEND results approach the resiliency of the *L3-follows-L1* upper-bound topology on high-impact failures at a substantially lower implementation cost.

The remainder of this paper is organized as follows. Section II reviews related work. Section III describes the ASCEND methodology, and Section IV presents experimental results on three markets. Section V discusses design choices, limitations, and future work, and Section VI concludes.

## II. Related Work

### A. Hierarchical and Core Network Design

Hierarchical organization has long been used to keep routing state and operational complexity manageable as networks grow [1], [3], [6], and core/edge separation is standard practice in service-provider metro design [4]. Selecting which sites should serve as core nodes has been studied as a location problem, for example via tabu search [11]. More broadly, survivable network design formulates topology selection under connectivity requirements as a combinatorial optimization problem, commonly solved with polyhedral and cutting-plane methods [12]. These approaches generally operate on a single layer: they select nodes and links but do not account for how logical links are realized over an existing fiber plant or for correlated physical failures.

### B. Survivable and Cross-Layer Routing

A complementary line of work takes the logical topology as given and asks how to route it over the physical layer so that it survives failures. Modiano and Narula-Tam showed that routing lightpaths so that the logical topology remains connected after any single fiber failure is NP-complete, and formulated the problem as an integer linear program [13]. Lee et al. generalized this to cross-layer connectivity metrics and showed that classical single-layer results such as the max-flow min-cut theorem no longer hold across layers [14]. Correlated failures are commonly modeled through SRLGs: computing SRLG-diverse paths is itself NP-complete [15], network availability changes substantially once SRLG failures are considered [7], and recent work develops efficient algorithms for region-disjoint survivable routing [16]. Survivable design against multiple failures has also been addressed by protecting spanning trees [17]. ASCEND's inner embedding step belongs to this family and uses a maximally resource-disjoint MCF formulation [18]. However, these methods take the logical topology, or the endpoint pairs, as input; they do not decide which sites become core sites or which core circuits to build.

### C. Metaheuristic Topology Optimization

Because topology design is combinatorial, heuristics and metaheuristics are widely used [19], [20]. Examples include the MENTOR heuristic for mesh topology design and routing [21], tabu search for topologies with unreliable components [22], genetic algorithms for distributed network topologies [23] and for reliability-driven design [24], multi-objective topology design [25], and heuristic optical topology optimization [26]. These approaches typically evaluate candidates at a single layer and under independent component failures, without embedding each candidate onto a physical fiber network.

### D. Positioning of This Work

ASCEND combines, within a single optimization loop, (i) core-site selection, (ii) core-circuit selection, (iii) a maximally disjoint, SRLG-aware embedding of every candidate onto fiber, (iv) evaluation of customer impact under correlated multi-failure scenarios, and (v) an explicit trade-off between that impact and hardware cost. To our knowledge, no prior approach addresses all five jointly.

## III. Method

### A. Overview

ASCEND is a systematic framework for Layer-3 network design that implements an automated optimization pipeline. At the Layer-3 level, ASCEND optimizes two primary design decisions: (1) which sites should function as core sites, and (2) how these core sites should be interconnected. Given these decisions, ASCEND then constructs the complete Layer-3 network (including access-to-core connectivity for edge sites) via the Layer-1 embedding step. This problem can be addressed under two different assumptions. In the first scenario, the set of core sites is fixed, and the outer optimization focuses solely on selecting the core circuits connecting these sites. In the second scenario, the core sites are not fixed; instead, ASCEND is provided with two sets of sites: (i) a set of mandatory core sites, which must remain part of the core, and (ii) a set of optional core sites, which may be promoted to core sites if deemed beneficial by the optimization. In this case, the outer optimization jointly determines the core-site selection and the core-circuit design.

Each candidate core design is uniquely represented by a vector encoding the selection of core sites and core circuits. Given this vector representation, two cost metrics—performance cost and implementation cost—are computed. As shown in Fig. 3, each candidate vector is pruned, decoded into a Layer-3 topology, mapped onto the physical Layer-1 network, and evaluated using performance and implementation cost functions. Using these cost functions, MOPSO searches for vector representations that yield resilient and cost-efficient Layer-3 network topologies.

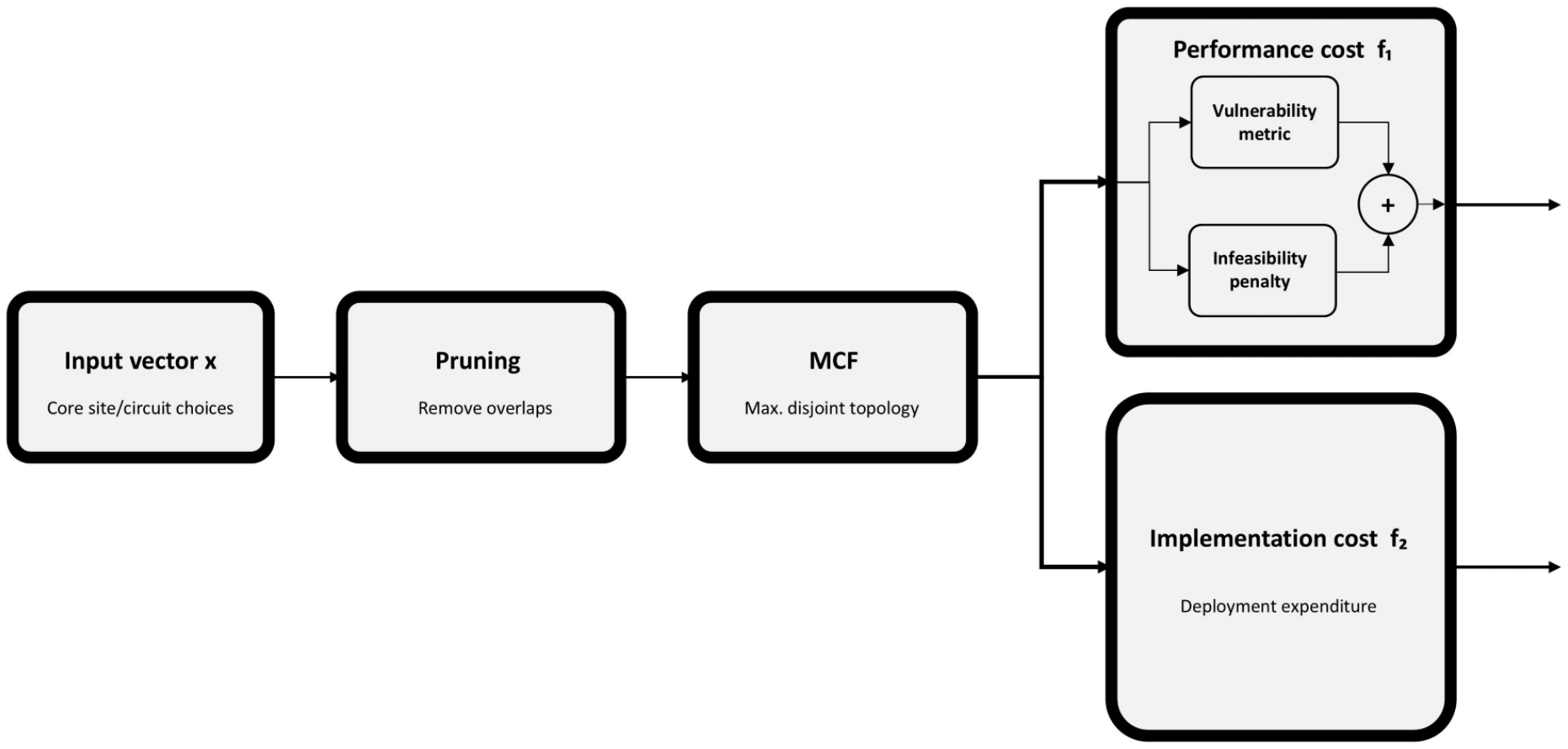

Fig. 3. How ASCEND converts a vector into a topology and evaluates cost functions.

### *B. Problem Formulation*

We now formally define the Layer-3 core design problem addressed in this paper, which is embedded to produce a complete Layer-3 network realization. Let the underlying physical network (Layer-1) be represented by a graph

$$G_1 = (V_1, E_1),$$

where $V_1$ denotes the set of sites and $E_1$ denotes the set of fiber links. Each fiber link $e \in E_1$ has an associated physical length, latency, and may belong to an SRLG. Let $S = \{S_1, S_2, \dots\}$ denote the set of all SRLGs, where each SRLG represents a correlated failure domain such that the failure of an SRLG simultaneously disables all links belonging to it.

A subset of sites $V_A \subseteq V_1$ are designated as ASBR (Autonomous System Boundary Router) sites that, in addition to serving customers, provide connectivity to the external or backbone network. Each site serves a known number of end customers (homes), which is used to quantify customer impact under failure scenarios.

Let $V_c \subseteq V_1$ denote the set of candidate core sites. A (logical) Layer-3 core network topology is defined by a graph

$$G_3 = (V_3, E_3),$$

where $V_3 \subseteq V_c$ is the selected set of core sites and $E_3 \subseteq V_3 \times V_3$ is the set of undirected core circuits interconnecting them. Each core circuit in $E_3$ must be realized as a path over the physical network $G_1$.

Given a candidate Layer-3 topology $G_3$, a physically feasible realization is obtained by routing all Layer-3 core circuits and access paths over $G_1$. This routing must respect physical topology constraints and should prioritize disjointness across physical links, nodes, and SRLGs while minimizing end-to-end latency.

The goal of the optimization is to jointly determine (i) which candidate sites should be selected as core sites, and (ii) how the selected core sites should be interconnected, so as to optimize multiple competing objectives under realistic failure assumptions.

Formally, the problem is posed as the following multi-objective optimization:

$$\min_{G_3} \begin{cases} \text{PerformanceCost}(G_3) \\ \text{ImplementationCost}(G_3) \end{cases}$$

Where $\text{PerformanceCost}(G_3)$ captures network resiliency and feasibility properties, and $\text{ImplementationCost}(G_3)$ represents the estimated hardware cost. Detailed definitions of these cost functions are provided in Section III-F.

### *C. Vector Representation*

Solving the problem requires representing each candidate design as a vector. Therefore, the first step is to define a suitable vector representation of a Layer-3 core network topology. Given $N$ candidate core sites, the Layer-3 core network is modeled as an undirected graph whose vertices correspond to candidate core sites and whose links correspond to potential core circuits between them. Each potential core circuit can be either present or absent. An adjacency matrix representation is used to encode core-site connectivity. Since the graph is undirected and self-loops are not allowed, only the upper (or equivalently, lower) triangular portion of the adjacency matrix is required to uniquely represent the circuits. As a result, the entire topology can be represented using $\frac{N(N-1)}{2}$ bits, where each bit indicates the presence or absence of a core circuit between a specific pair of candidate core sites.

Interpretation of the connectivity bits can be done as follows:

- $e_{ij} = 1$ if there is a circuit between sites $i$ and $j$.
- $e_{ij} = 0$ otherwise.

Such bits can form a vector such as:

$$vector = [\, e_{12}, e_{13}, \dots, e_{1N} | e_{23}, e_{24}, \dots, e_{2N} | \dots | e_{N-1N}]$$

Hence, the full vector length, $len$, is:

$$len = \frac{N(N-1)}{2}$$

For example, if 4 possible core sites, such as $\{A, B, C, D\}$, exist, the vector will be:

$$vector = [\, e_{AB}, e_{AC}, e_{AD}, e_{BC}, e_{BD}, e_{CD}]$$

Core-site selection is represented implicitly through connectivity: a site is considered a core site if and only if it participates in at least one selected core circuit. Sites with zero incident core circuits are treated as edge sites. This implicit encoding avoids the need for additional node-selection variables and yields a compact representation in which both core-site selection and core-circuit selection are determined jointly by the connectivity vector.

*D. Pruning*

When the Layer-3 circuits are mapped to Layer-1 infrastructure, some core circuits may pass through intermediate core sites. For example, Fig. 4 illustrates a hypothetical $G_1$ section with three core sites. Considering all possible node pairs, this results in three core circuits: $[\, e_{AB}, e_{AC}, e_{BC}]$. However, when the circuit $e_{AC}$ is mapped to $G_1$, it traverses core site $B$. Under the circuit and routing model used in this work, an $A$-to-$C$ core circuit whose physical path passes through intermediate core site $B$ is treated as redundant with respect to the retained $A$-$B$ and $B$-$C$ connectivity. We therefore remove such circuits to reduce the decision space. In the example above, the resulting pruned core circuits are $[\, e_{AB}, e_{BC}]$.

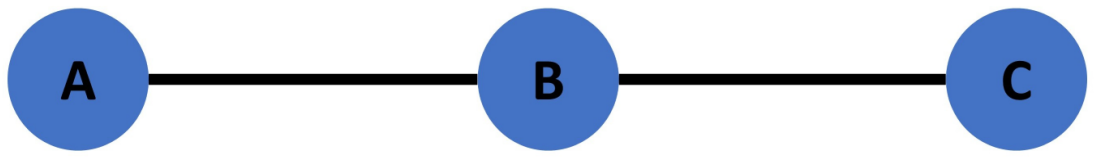

Fig. 4. A hypothetical $G_1$ section with 3 core sites.

When ASCEND is executed with fixed core sites, the outer optimization varies only the selection of core circuits. As a result, redundant circuits remain unchanged across optimization runs. In this scenario, pruning can be applied prior to the outer optimization to reduce the size of the search space by shortening the decision vector. Assuming that the vector length, $len$, is decreased to $len' = len/p$, where $p$ is the pruning factor, since the vector is binary (each circuit can exist or not exist), the size of the search reduces from $2^{len}$ to $2^{len/p}$, which is equal to $\sqrt[p]{2^{len}}$. In our practical experiments for runs with fixed core sites, $p$ usually has a value close to 2, meaning that half of the circuits are usually pruned. This means that the magnitude of the search space is reduced to the square root of the original value, effectively making the problem easier to solve.

*E. Inner Routing/Embedding via Maximally-Disjoint Integral MCF*

After pruning, we obtain a set of core sites and core circuits that must be translated into a complete network topology, including the selection of edge circuits and the mapping of Layer-3 circuits onto Layer-1 fiber paths (see Fig. 1). This process determines a fiber-path realization for each core circuit, selects the edge circuits, and determines the corresponding Layer-1 paths for all edge circuits. To achieve a resilient implementation, the mapping must ensure that at Layer-1, the choice of links and nodes is as disjoint as possible. Technically, the mapping must satisfy the following criteria:

- Maximum SRLG disjointness
- Maximum link disjointness
- Maximum node disjointness
- Lowest delay

We use our internally developed integral MCF formulation [18] for routing that explicitly minimizes resource overlap across links, nodes, and SRLGs by penalizing each additional reuse of a resource beyond an allowed baseline with escalating costs, ensuring overlap minimization takes precedence over traditional path cost objectives while still minimizing delay. At a high level, the MCF step is an integral embedding that jointly selects Layer-1 paths for all circuits, subject to standard feasibility constraints (e.g., flow conservation and capacity). Disjointness is enforced in a "maximally disjoint" sense by counting and penalizing repeated use of physical resources (links, nodes, and SRLG risk groups) beyond a small baseline allowance; the penalty weights are chosen so that reducing overlap is prioritized before reducing delay.

Given the choice of core sites and core circuits, MCF outputs the resulting embedding (i.e., the set of Layer-1 paths realizing the full Layer-3 design). As a result, for each candidate vector evaluated by MOPSO, the MCF formulation optimally solves the lower-level embedding problem for the fixed candidate design (mapping $G_3$ to $G_1$) under the specified disjointness-first objective. Full formulation details are provided in a companion manuscript currently in preparation [18].

### *F. Cost Functions*

Given a network topology, two cost functions must be minimized: performance cost (how bad the performance issues are) and implementation cost (how much money is needed to implement this topology).

#### *1) Performance Cost*

Performance cost measures the inefficiency of a topology, $G_3$, that we aim to minimize. It has 2 components:

1. **Vulnerability**: How prone this network is to faults. In other words, in case of a fault, how many homes get disconnected?
2. **Infeasibility penalty**: Some topologies can have core-connectivity that is technically infeasible. Also, if certain sites are required to be core, absence of them as core sites is undesirable. A penalty is added for existence of such cases in the topology.

The total value of performance cost is calculated as the sum of these 2 components, explained in the rest of this section.

*a) Vulnerability Metric*

The vulnerability metric quantifies how susceptible a network topology is to disruptions. Our goal is to make the topology as resilient as possible, which translates to minimizing this metric. Since fiber cuts can occur at different locations, we test multiple cut locations and measure how many homes become isolated (disconnected) in each fiber cut scenario. The topology's total vulnerability metric is then computed based on the number of disconnected homes across all scenarios. In other words, the vulnerability metric is calculated in two steps: (1) isolation analysis, which identifies the nodes and homes disconnected in each failure scenario, and (2) penalty calculation, which assigns a penalty to each fault scenario and sums the penalty of all scenarios, scaled by their probability.

*i) Isolation Analysis*

The goal of this analysis is to quantify how many homes become disconnected under each fault scenario. The resiliency objectives of the proposed architecture can be expressed in terms of its ability to maintain connectivity in case of $X$ concurrent failures (fiber cuts). The value of $X$ can be chosen according to the resiliency requirements and risk tolerance of the network operator. In many service provider networks, single-failure scenarios ($X = 1$) rarely result in customer isolation because alternative paths are typically available. As the number of simultaneous failures increases, however, the likelihood of isolated nodes or disconnected regions grows. While large values of $X$ correspond to highly improbable events and are therefore of limited practical interest, moderate values provide a useful lens for evaluating network robustness.

Fiber-cut risk can be modeled using different levels of detail depending on the availability of data. When historical fiber-cut records or route-specific characteristics are available, such as aerial versus underground deployment, environmental conditions, construction exposure, or other vulnerability indicators, segment-level risk may be estimated using a weighted-exposure approach that incorporates these factors. In the absence of such information, a simpler approximation may be used in which cut likelihood is assumed to be proportional to exposed fiber length, corresponding to a uniform hazard rate per unit distance and representing a first-order approximation of a Poisson hazard model. Under this approximation, we assume initiating failure locations are independent except where correlation is modeled through SRLG membership, consistent with probabilistic SRLG frameworks that explicitly represent correlated failure domains. The model is intended as a design-stage risk proxy for comparing and ranking network topologies rather than as an absolute predictor of outage frequency.

After choosing the value of $X$, we consider all possible combinations of $X$ faults. Because a fault may occur either within an SRLG region or along a non-SRLG fiber segment, we consider both cases:

**I. Non-SRLG cut**

In this case, the fiber is cut on the non-SRLG section. To calculate the probability of the cut, we first determine the length of that non-SRLG portion. For fibers that do not belong to any SRLG group, the non-SRLG portion is simply the total fiber length. For links that do belong to an SRLG group, the desired non-SRLG length ($\hat{l}$) is the link length ($l$) minus the total length of the link that falls within SRLGs ($l_{SRLG}$). In mathematical terms:

$$\hat{l} = \begin{cases} l & if\ l\ does\ not\ belong\ to\ any\ SRLG\ group \\ l - l_{SRLG} & if\ l\ belongs\ to\ some\ SRLG\ group \end{cases}$$

Now the probability of this cut, called $c$, is:

$$Prob(c) = \frac{\hat{l}}{l_{total}}$$

Where $l_{total}$ is the total fiber length in the network.

**II. SRLG cut**

In this event, the fiber is cut on the SRLG section. When the SRLG group goes down, all links in that group are disabled. Based on the assumption that probability of a fiber cut event is proportional to the fiber length, the probability of a fiber cut in a certain SRLG region is equal to the length of the SRLG ($l_{SRLG}$), divided by the total fiber length in the market ($l_{total}$).

$$Prob(c) = \frac{l_{SRLG}}{l_{total}}$$

Given an event consisting of $X$ cuts, where each cut may occur either within an SRLG region or outside any SRLG region, the total probability of the event is:

$$Prob(event) = \prod_{k=1}^{X} Prob(c_k)$$

Note that we have already addressed the dependence of the fiber cut events in SRLGs. After representing each SRLG as a single correlated failure domain, we assume independence among distinct initiating failure domains. Under this modeling assumption, the unnormalized weight of an $X$-failure scenario is computed as the product of its constituent failure-domain weights. These scenario weights are used to rank candidate topologies consistently and are not intended as calibrated estimates of absolute outage probabilities.

After removing the disconnected links in each scenario, we analyze the remaining graph to identify the nodes that are no longer connected to the backbone due to the fiber cuts. For each disconnected node, we retrieve the number of homes served by that node from our dataset, which allows us to compute the number of customers disconnected under each fault scenario.

*ii) Penalty Calculation*

For each outage, the vulnerability must increase with both the outage size (i.e., the number of homes disconnected) and the probability of the event.

Since outages that disconnect more homes are more undesirable, the penalty for each fault event is designed to increase exponentially with the number of isolated homes. Formally, the penalty ($Penalty$) of each outage is calculated as follows:

$$Penalty\left(N_{DC}(event)\right) = b^{(N_{DC}/c)} - 1$$

where $b > 1$ is the exponential base (set to 1.1 in our analysis, see Section V-C), $c$ is a scaling factor (set to $10000$) to keep the metric within a numerically manageable range, and $N_{DC}(event)$ is the number of homes disconnected in the $event$. The exponential function was used to assign a higher penalty to large outages. The term $-1$ is used to make sure that the penalty is 0 if no home gets disconnected. Fig. 5 shows the penalty that this formula assigns for each number of disconnected homes.

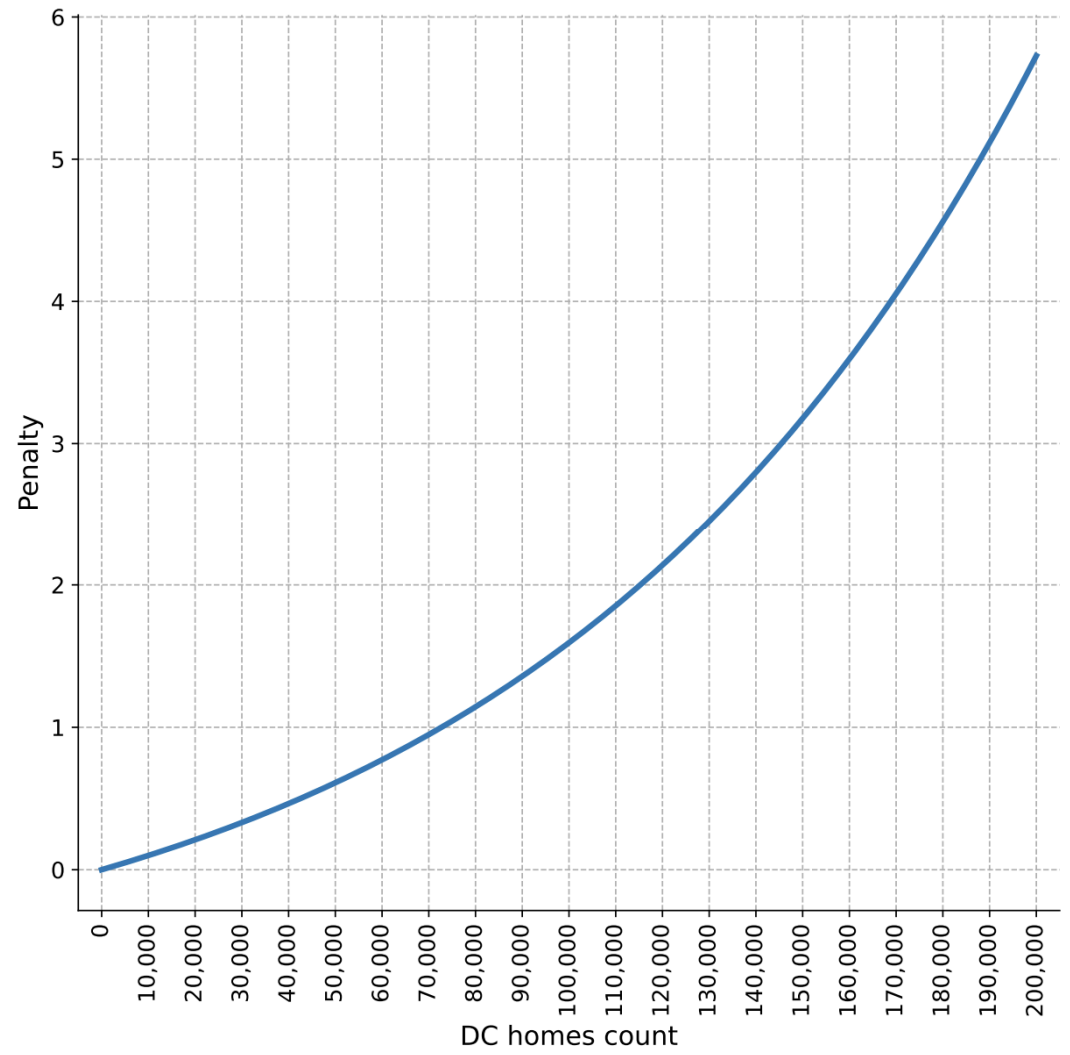

Fig. 5. Exponential penalty as a function of the number of disconnected homes.

The total vulnerability of the network is the sum of the penalties of all outage events multiplied by their probability:

$$\text{Vulnerability}(G_3) = \sum_{event \in AllEvents} Prob(event) \times Penalty\left(N_{DC}(event)\right)$$

where $Prob(event)$ is the probability of each outage event (described in Section III-F1a-i), $N_{DC}(event)$ is the number of disconnected homes, and $Penalty$ is the penalty corresponding to the number of homes disconnected in the outage event. After calculating the penalty of each outage event, and summing over all outage events (described in Section III-F1a-i), the total vulnerability of the network design is calculated.

*b) Infeasibility Penalty*

There may be some architectural restrictions in designing a network for various reasons. For example, the number of core connections for each core site must remain within specified minimum and maximum bounds. In order to avoid the infeasible topologies, the performance cost function must be higher if the design is infeasible. This is regardless of the vulnerability of the design. To avoid these issues, we add another component to the objective function, called the infeasibility penalty.

If we look at the core-connectivity of the Layer-3 graph, and denote the core degrees of the core site $i$ with $d_i$, and define $min_{d_i}$ as the minimum acceptable degree, $max_{d_i}$ as the maximum acceptable degree for site $i$, and $missing_j$ is set to 1 if site $j$ is a required core site but is an edge site, the infeasibility penalty of the design will be:

$$\text{Infeasible}(G_3) = \lambda_1 \sum_{i \in AllCore} (d_i < min_{d_i}) + \lambda_2 \sum_{i \in AllCore} (d_i > max_{d_i}) + \lambda_3 \sum_{j \in RequiredCore} missing_j$$

where $\lambda_1, \lambda_2, \lambda_3$ are scaling factors. Expressions such as $(d_i < min_{d_i})$ evaluate to 1 if true and 0 otherwise.

Then, the resulting performance cost is:

$$\text{PerformanceCost}(G_3) = \text{Vulnerability}(G_3) + \text{Infeasible}(G_3)$$

In practice, we want to ensure that all final output topologies avoid all infeasibility issues ($\text{Infeasible}(G_3) = 0$). This can be achieved by setting each $\lambda_k$ to a value that is larger than practical values of $\text{Vulnerability}(G_3)$. In our implementations, we set all $\lambda_k > 1$, which ensures that any topology with $\text{PerformanceCost}(G_3) < 1$ has 0 infeasibility issues. The value of 1 was chosen because it is much larger than the practical values of $\text{Vulnerability}(G_3)$ for good topologies. Although this threshold was empirically chosen, we found the framework to be robust to moderate variation in $\text{Vulnerability}(G_3)$ and $\text{Infeasible}(G_3)$, as long as infeasibility penalties dominate feasible vulnerability values.

*2) Implementation Cost and Capacity Forecasting*

Implementation cost is estimated using an internal planning model that evaluates the resources required to support the projected traffic and resiliency requirements of each candidate topology. The model considers topology-dependent capacity requirements under representative operating and failure conditions and produces an aggregate implementation-cost estimate. Implementation-specific forecasting, equipment-sizing, architectural, and pricing details are omitted to protect proprietary and business-sensitive information.

For the comparative evaluation, the same cost-estimation methodology is applied consistently to all candidate and benchmark topologies. The resulting costs are normalized as described in Section IV to avoid disclosing absolute cost values.

*G. Metaheuristic Search*

Now that each topology can be represented as a design vector and evaluated through the objective functions, the problem reduces to identifying the optimal vector. The size of the design space, however, grows combinatorially with the number of candidate sites. The total number of possible topologies scales as $2^{\frac{N(N-1)}{2}}$, which grows exponentially with the square of the number of candidate core sites. As a result, exhaustive enumeration (brute-force search) becomes computationally infeasible, even for relatively small network sizes (see Appendix A for more details about the search magnitude). This combinatorial explosion motivates the use of intelligent optimization techniques, which can effectively explore the most promising regions of the search space without requiring complete enumeration.

*1) Search Algorithm: MOPSO*

Metaheuristics are well suited to this problem because the number of binary design decisions grows quadratically with the number of candidate sites, and exact formulations become computationally prohibitive at realistic scales [12].

ASCEND is based on Particle Swarm Optimization (PSO) [8] for optimizing the topology. PSO is a population-based optimization algorithm inspired by the social behavior of birds flocking or fish schooling. In PSO, a group of candidate solutions—called particles—move through the solution space, adjusting their positions based on their own experience and that of their neighbors. Each particle tracks its best-known position and is attracted toward the global best position found by the swarm. Given the binary nature of our problem, the binary version of PSO was implemented [10]. Also, since we want to minimize two cost functions, namely performance cost and implementation cost, ASCEND uses MOPSO [9], which is an extension of the standard PSO designed to solve problems involving multiple conflicting objectives (see Fig. 6). We selected MOPSO after testing different metaheuristic search algorithms (see Section V-A).

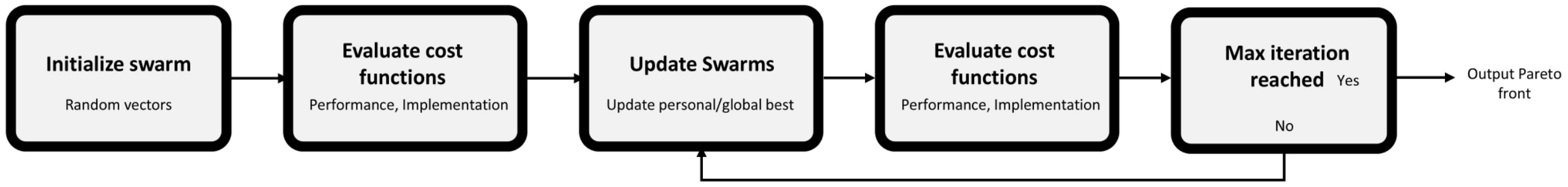

Fig. 6. MOPSO steps.

*2) Pareto Front*

In multi-objective optimization, solutions are evaluated with respect to multiple, often conflicting objectives. A solution is said to be Pareto-optimal if no other feasible solution is at least as good in all objectives and strictly better in at least one objective. Equivalently, a Pareto-optimal solution cannot be improved in one objective without degrading at least one other objective. The set of all Pareto-optimal solutions forms the Pareto front, which represents the best achievable trade-offs among the objectives in the objective space. Rather than producing a single optimal solution, multi-objective optimization algorithms generally aim to approximate the Pareto front, thereby providing decision makers with a spectrum of alternative solutions reflecting different compromises among the objectives [27]. After obtaining an approximation of the Pareto front, a decision maker can select, for example, a more resilient design at a higher implementation cost or a less resilient design at a lower implementation cost.

## IV. Results

### *A. Experimental Setup*

This section first describes the experimental setup and then reports the results of applying ASCEND to three representative Layer-3 networks derived from an existing Layer-1 graph in a service provider network in the United States. The presented network instances and results are intended solely to demonstrate the methodology and associated trade-offs and should not be construed as representing the architecture, configuration, or operational characteristics of any actual service-provider network. All data have been de-identified to protect proprietary and business-sensitive information.

To prevent disclosure of absolute cost values, all implementation costs are normalized by the implementation cost of the corresponding *L3-follows-L1* reference topology, described in Section IV-A2b. Consequently, the normalized cost of *L3-follows-L1* is 1.0 for each market. We report optimization results up to the Pareto front. After that, the final network topology can be chosen from the Pareto front candidates based on various business needs.

After running ASCEND and generating the full set of candidate solutions, we restrict our analysis to a practical region of interest in which both performance cost and implementation cost fall below predefined thresholds. Solutions that achieve very low values in one objective at the expense of excessively high values in the other objective are excluded, as such trade-offs are impractical for real-world deployment. For completeness, the full set of discovered topologies is provided in Appendix B. An important characteristic of the selected region is that all solutions exhibit 0 infeasibility penalty, and the performance cost is identical to the vulnerability metric (see Section III-F1b).

Because MOPSO is stochastic, we report results for a single random seed throughout the main experiments for consistency. To verify the effect of the stochastic factor, we ran ASCEND using independent seeds and show the resulting Pareto fronts in Appendix C.

*1) Hardware and Run Time*

All experiments were run on an Intel® Xeon® Platinum 8280–based system (2.70 GHz) with 16 CPU threads. We parallelized the pipeline based on the number of available processors. Data for different markets are shown in Table I. Sample size and maximum number of iterations were adjusted based on market size and the evaluation time per topology, to achieve a reasonable balance between solution quality and run time.

TABLE I
MARKET DATA AND RUN DETAILS.

| Market | # of core site candidates | # of required core sites | Sample size | Max iterations | Wall clock |
|---|---|---|---|---|---|
| **Market A** | 13 | 10 | 2000 | 50 | 2D, 1H, 32M |
| **Market B** | 8 | 6 | 8000 | 60 | 6D, 6H, 41M |
| **Market C** | 9 | 9 | 2000 | 50 | 7D, 2H, 55M |

*2) Comparison Benchmarks*

We compare the results of ASCEND against two benchmarks: baseline and *L3-follows-L1*, explained in what follows.

*a) Baseline*

Before implementing ASCEND, we used a semi-automated process to design a resilient network. This approach consisted of several automated steps, with manual review and modification performed between stages. We use the topologies generated by this pipeline as the baseline and compare them against the results produced by ASCEND. For technical details of the baseline topology, refer to Appendix D.

*b) L3-follows-L1*

In addition to the manually designed baseline, we compare the results of ASCEND against the most resilient Layer-3 network that can be constructed for a given Layer-1 infrastructure. In this design, every Layer-1 site is promoted to a Layer-3 core site, and every Layer-1 link is instantiated as a Layer-3 core circuit. We refer to this topology as *L3-follows-L1*. While this approach achieves the maximum possible resiliency and performance, it does so at an extremely high implementation cost, particularly when the number of nodes and edges is large, rendering it impractical in real deployments and useful only as a theoretical performance ceiling. For each market, this topology is used for normalizing implementation costs.

## B. Market A

Market A is a relatively small market with 13 candidate core sites. We ran ASCEND with a sample size of 2000 and 50 iterations. After running the MOPSO pipeline, we visualized solutions with low performance cost and low implementation cost (Fig. 7). Each point represents a network design plotted by performance cost and implementation cost. The manual baseline design is marked by a black circle. For comparison, the limits of the manual baseline are shown as dashed black lines. ASCEND identified multiple solutions that improve both performance cost and normalized implementation cost relative to the baseline, as well as additional nondominated solutions representing alternative trade-offs. Pareto-front solutions are indicated by diamond markers in various colors. Note that the implementation costs are normalized to avoid disclosing sensitive information.

Within the low-cost region of interest, 8 Pareto-front solutions were identified, for which a more detailed vulnerability analysis is provided in Table II, where the numbers of isolated homes across failure scenarios are shown. To provide a more intuitive interpretation, we clustered failure scenarios into groups based on the number of homes they isolate. Group 0 represents the number of failure scenarios that result in no homes being isolated. The next group includes failure scenarios that isolate between 1 and $H$ homes, and subsequent groups include scenarios that isolate between $H+1$ and $2H$ homes, and so on. We do not disclose the value of $H$ to protect sensitive information. A more resilient design is characterized by having as few failures as possible in the higher-isolation groups. We call this representation the failure profile. It should be noted that failure profile alone does not show the full vulnerability metric, because it merely reflects the counts of failures in each group regardless of the probability of the failure event, or where the events fall within the group. The failure profiles for the manual baseline and *L3-follows-L1* are shown in the two rightmost columns for comparison.

Based on the results of Table II, topologies 46761 and 64729 achieve the same failure profile as the *L3-follows-L1* configuration—which represents the theoretical performance limit—across the three higher-impact failure groups. Through joint optimization of performance and cost, these topologies exhibit strong shockproof behavior while maintaining significantly lower implementation costs.

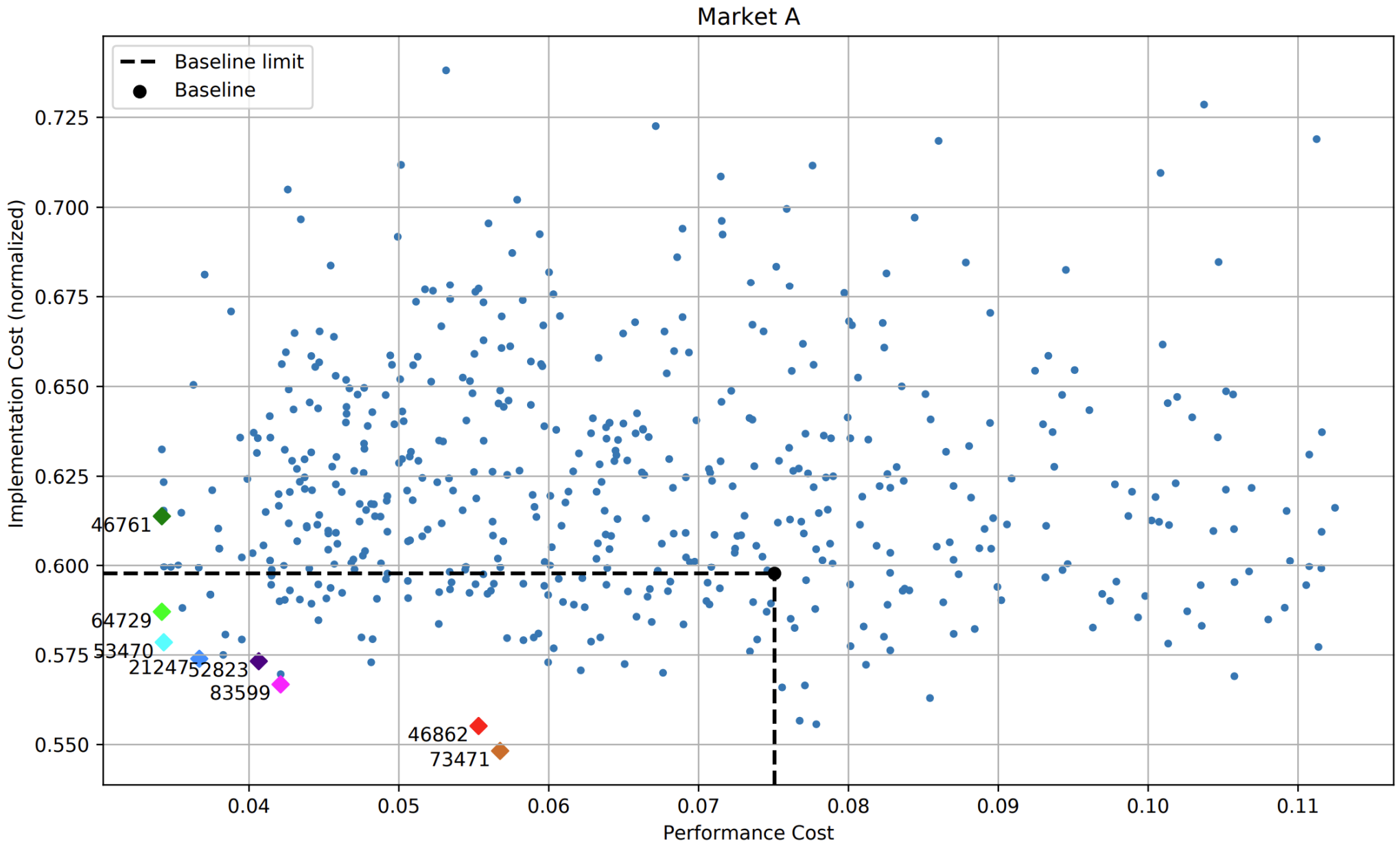

Fig. 7. Market A. Visualization of solutions with low performance cost and low implementation cost.

TABLE II
MARKET A. ANALYSIS OF THE PARETO FRONT SOLUTIONS.

| **Group** | **Start** | **End** | **Failure Event Count** | | | | | | | | | |
|---|---|---|---|---|---|---|---|---|---|---|---|---|
| | | | **Topology 46761** | **Topology 64729** | **Topology 53470** | **Topology 21247** | **Topology 52823** | **Topology 83599** | **Topology 46862** | **Topology 73471** | **Baseline** | ***L3-follows-L1*** |
| 0 | 0 | 0 | 3220 | 3219 | 3220 | 3204 | 3180 | 3187 | 3182 | 3128 | 3261 | 3384 |
| 1 | 1 | $H$ | 474 | 474 | 471 | 477 | 475 | 481 | 476 | 464 | 465 | 444 |
| 2 | $H+1$ | $2H$ | 333 | 334 | 334 | 333 | 354 | 345 | 342 | 338 | 274 | 216 |
| 3 | $2H+1$ | $3H$ | 40 | 40 | 40 | 48 | 47 | 51 | 52 | 41 | 48 | 23 |
| 4 | $3H+1$ | $4H$ | 21 | 21 | 22 | 22 | 22 | 22 | 26 | 102 | 30 | 21 |
| 5 | $4H+1$ | $5H$ | 6 | 6 | 7 | 7 | 8 | 8 | 10 | 17 | 15 | 6 |
| 6 | $5H+1$ | $6H$ | 1 | 1 | 1 | 3 | 7 | 1 | 5 | 4 | 1 | 1 |
| 7 | $6H+1$ | $7H$ | 0 | 0 | 0 | 1 | 2 | 0 | 2 | 1 | 1 | 0 |

| | | | | | | | | | | |
|---|---|---|---|---|---|---|---|---|---|---|
| **Vulnerability Metric** | 0.0342 | 0.0342 | 0.0343 | 0.0367 | 0.0406 | 0.0421 | 0.0553 | 0.0568 | 0.0751 | 0.0189 |
| **Implementation Cost (normalized)** | 0.6139 | 0.5871 | 0.5785 | 0.5739 | 0.5733 | 0.5668 | 0.5552 | 0.5482 | 0.5978 | 1.0000 |

### *1) Analyzing One Failure Scenario*

Based on the mapping between Layer-1 and Layer-3, a fiber cut at Layer-1 can translate into one or more circuit cuts at Layer-3 (see Section III-B). When multiple fiber cuts belong to different SRLGs, multiple Layer-1 links are disconnected, which in turn results in a severe shock caused by the disconnection of multiple Layer-3 circuits. This behavior can be observed in Table II, where the baseline solution exhibits a failure event clustered in Group 7.

We analyze this specific failure scenario in more detail. Multiple fiber cuts may cause multiple SRLGs to fail, which translates into multiple circuits going down at Layer-3. As a result, three core sites become isolated from the backbone, which subsequently isolates five additional edge sites. More than $6H$ homes are disconnected (Fig. 8), making this an extremely undesirable outcome.

In contrast, in Solution 53470, ASCEND has added one core site and arranged core circuits to avoid such large-scale isolation. Under the same fiber cut scenario, no homes are disconnected (Fig. 9). In addition to improving resiliency, Solution 53470 has a lower normalized implementation cost than the baseline, partly due to more efficient device placement.

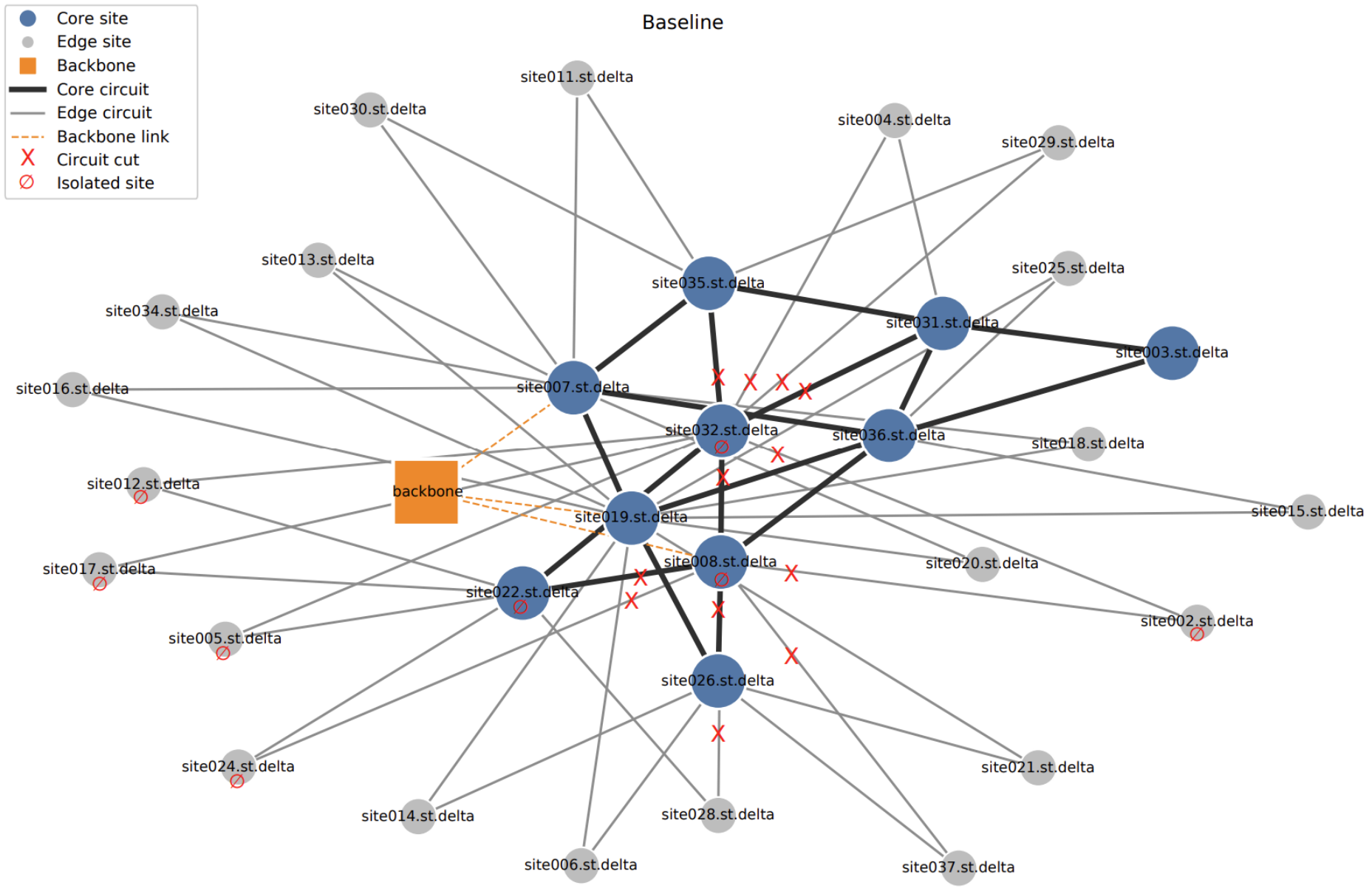

Fig. 8. Market A. Baseline network. One event will isolate 8 sites.

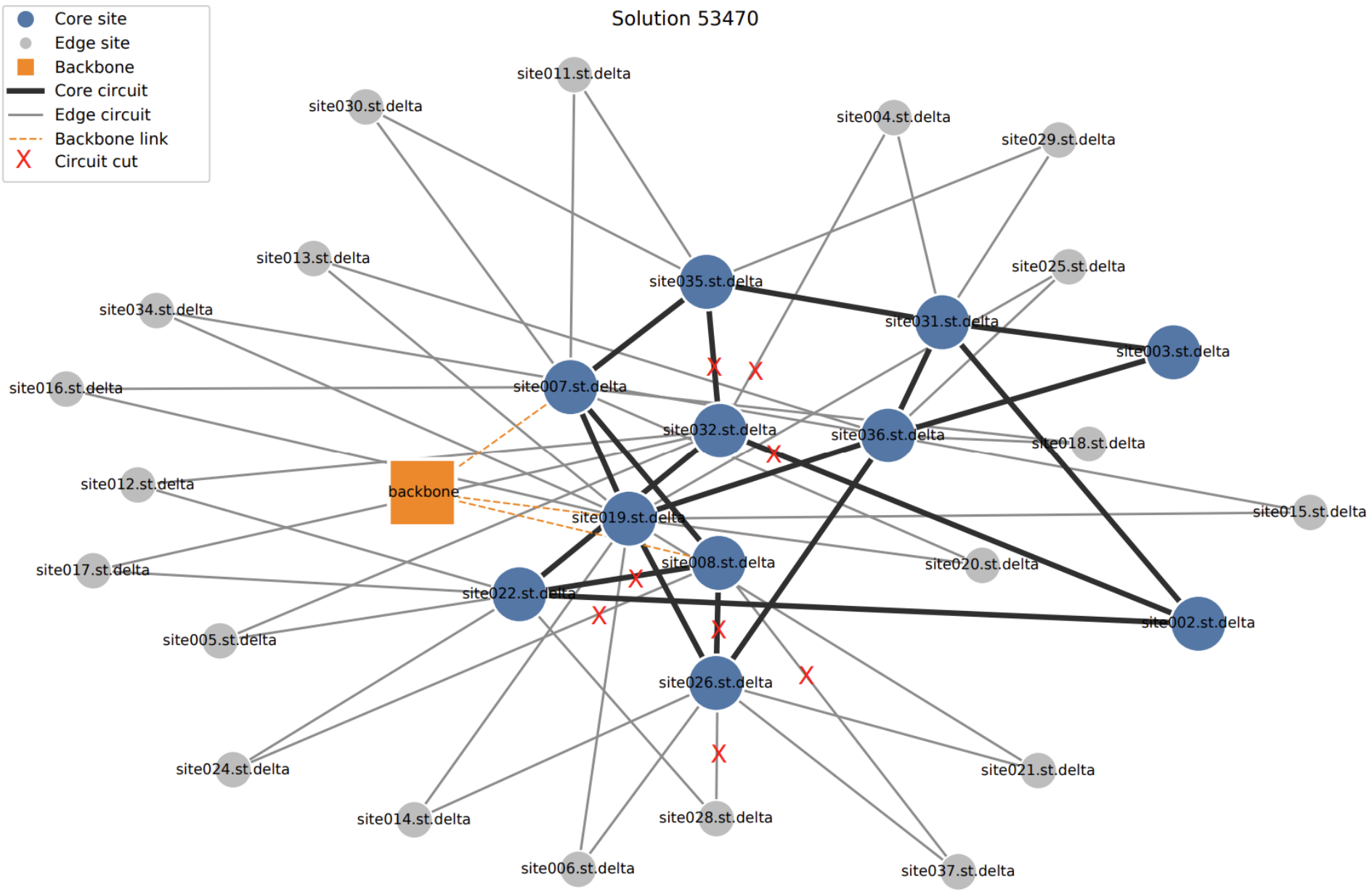

Fig. 9. Market A. Solution 53470. No sites or homes will be isolated.

## C. Market B

Market B includes eight candidate core sites, of which six must be designated as core. Failure profiles for selected Pareto-front topologies (annotated in Fig. 10) are summarized in TABLE III. To keep the presentation concise, only a subset of sufficiently distinct Pareto solutions is shown. Most Pareto-front solutions eliminate the group 4 failure present in the baseline design. In addition, many solutions reduce group 3 failures from 11 in the baseline to 4, approaching the theoretical minimum of 2 observed in the *L3-follows-L1* configuration. Although the manually designed baseline achieves a very low cost, ASCEND identifies several solutions that improve both objectives relative to the baseline, while other nondominated solutions offer lower implementation cost in exchange for higher performance cost.

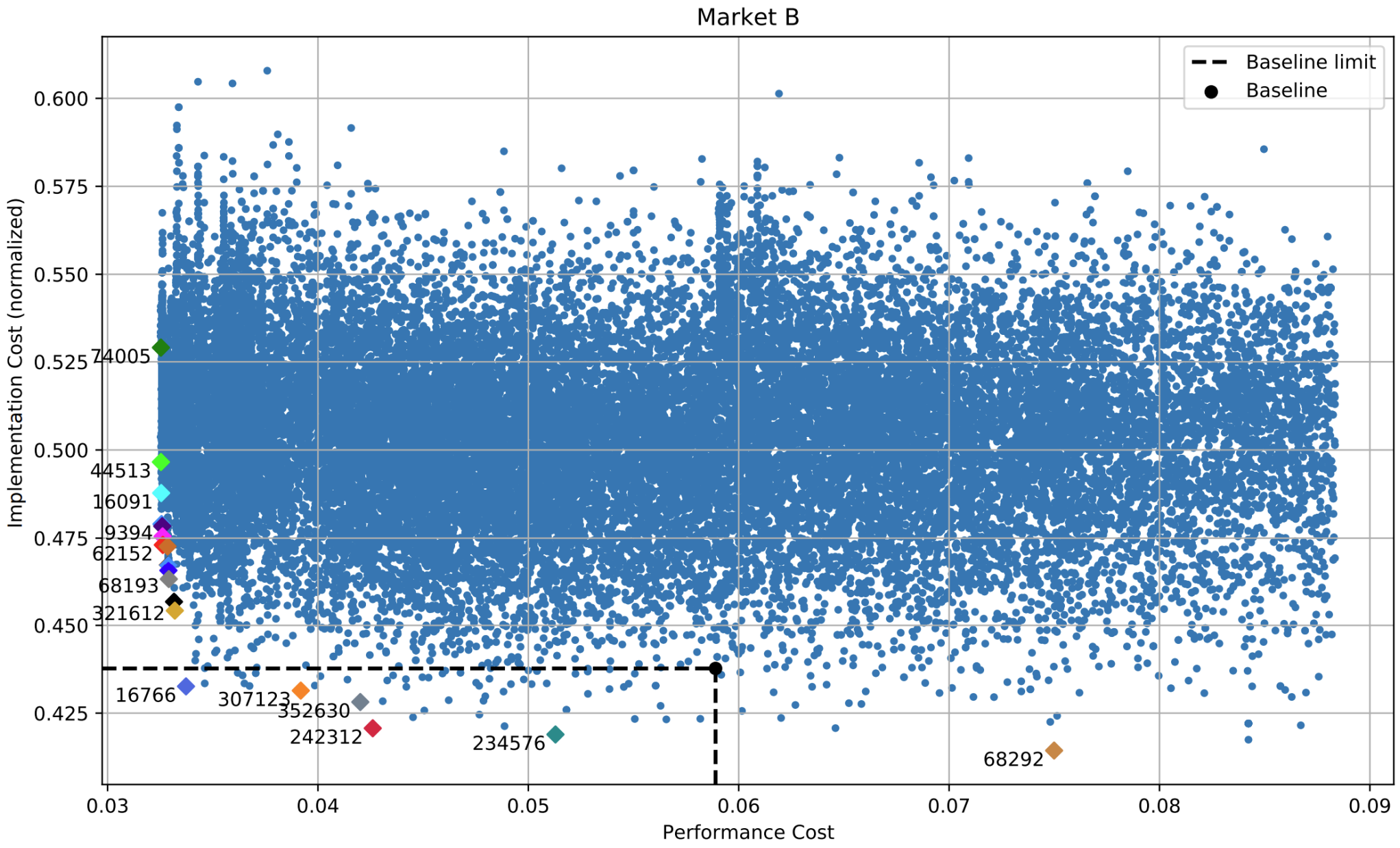

Fig. 10. Market B. Visualization of solutions with low performance cost and low implementation cost.

TABLE III
MARKET B. ANALYSIS OF THE PARETO FRONT SOLUTIONS.

| Group | Start | End | Failure Event Count | | | | | | | | | | | | | | |
|---|---|---|---|---|---|---|---|---|---|---|---|---|---|---|---|
| | | | Topology 74005 | Topology 44513 | Topology 16091 | Topology 9394 | Topology 62152 | Topology 68193 | Topology 321612 | Topology 16766 | Topology 307123 | Topology 352630 | Topology 242312 | Topology 234576 | Topology 68292 | Baseline | *L3-follows-L1* |
| 0 | 0 | 0 | 2009 | 2007 | 2006 | 2001 | 1999 | 1989 | 1986 | 1964 | 1923 | 1943 | 1871 | 1887 | 1818 | 1832 | 2181 |
| 1 | 1 | $H$ | 728 | 729 | 729 | 730 | 730 | 733 | 741 | 760 | 775 | 758 | 824 | 806 | 784 | 834 | 582 |
| 2 | $H+1$ | $2H$ | 33 | 34 | 35 | 39 | 41 | 45 | 41 | 46 | 64 | 61 | 64 | 75 | 158 | 96 | 9 |
| 3 | $2H+1$ | $3H$ | 4 | 4 | 4 | 4 | 4 | 5 | 4 | 4 | 12 | 10 | 9 | 6 | 8 | 11 | 2 |
| 4 | $3H+1$ | $4H$ | 0 | 0 | 0 | 0 | 0 | 2 | 2 | 0 | 0 | 2 | 5 | 0 | 6 | 1 | 0 |
| 5 | $4H+1$ | $5H$ | 0 | 0 | 0 | 0 | 0 | 0 | 0 | 0 | 0 | 0 | 1 | 0 | 0 | 0 | 0 |

| | | | | | | | | | | | | | | | |
|---|---|---|---|---|---|---|---|---|---|---|---|---|---|---|---|
| Vulnerability Metric | 0.0325 | 0.0325 | 0.0325 | 0.0326 | 0.0326 | 0.0329 | 0.0332 | 0.0337 | 0.0392 | 0.0420 | 0.0426 | 0.0513 | 0.0750 | 0.0589 | 0.0178 |
| Implementation Cost (normalized) | 0.5290 | 0.4966 | 0.4878 | 0.4790 | 0.4731 | 0.4632 | 0.4542 | 0.4326 | 0.4314 | 0.4282 | 0.4207 | 0.4189 | 0.4143 | 0.4377 | 1.0000 |

### *D. Market C*

Market C is a relatively large market. To reduce the computational burden, ASCEND was executed with fixed core site choices, sample size of 2000 and 50 iterations, so that the run is finished within a reasonable time period. Fig. 11 shows a scatter plot of the candidate solutions and Table IV shows the failure profile of the Pareto-front topologies. For most Pareto-front solutions, the numbers of high-impact failure scenarios are substantially lower than those of the baseline: The number of Group 5 failure scenarios decreased from 5 to 2, and the number of Group 4 failures decreased from 11 to 2. The failure counts in the two highest-impact groups equal those of *L3-follows-L1*, while the implementation costs are less than half as large. The results indicate that ASCEND produces multiple solutions that outperform the baseline in both objectives, together with additional nondominated trade-off solutions.

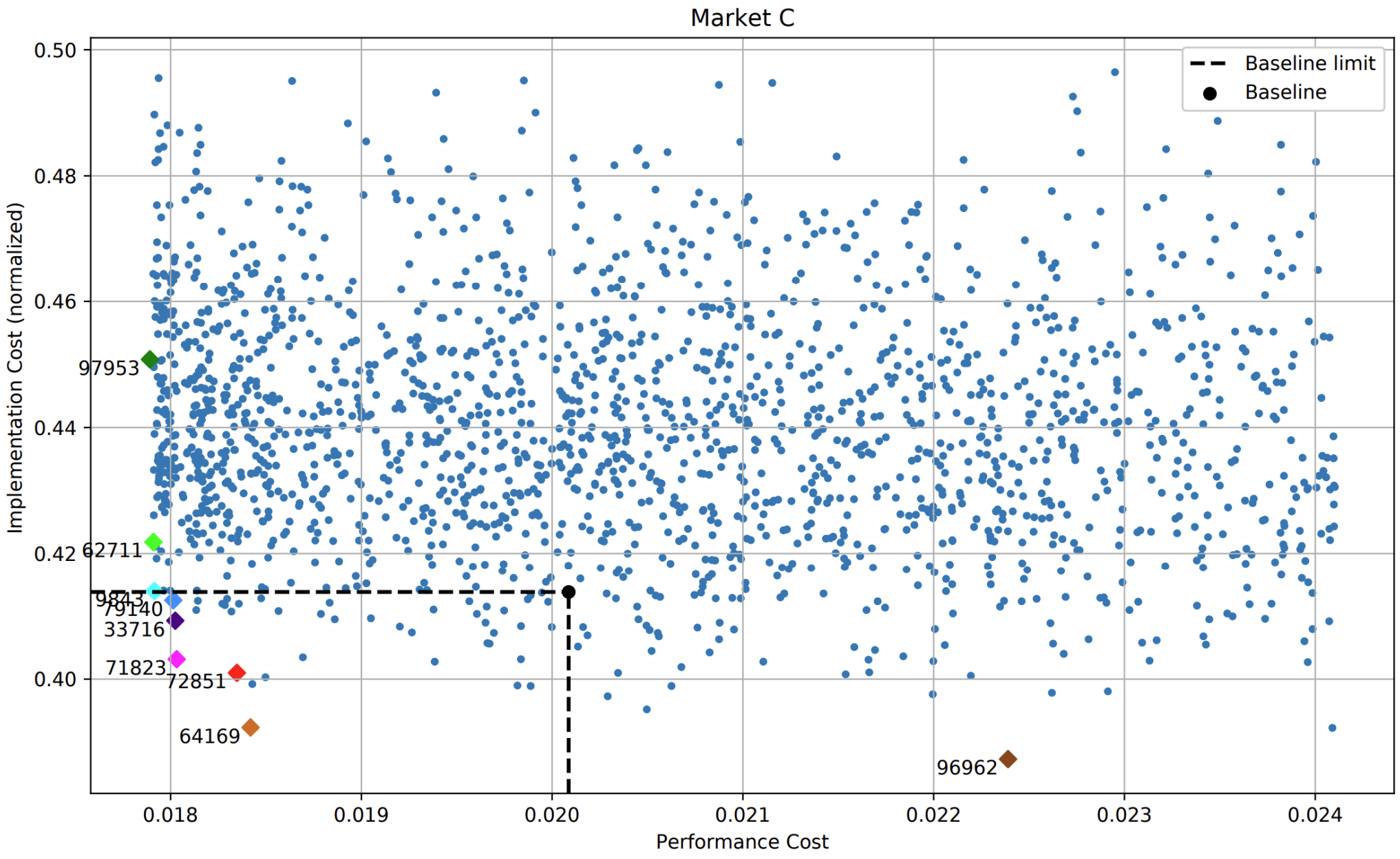

Fig. 11. Market C. Visualization of solutions with low performance cost and low implementation cost.

TABLE IV
MARKET C. ANALYSIS OF THE PARETO FRONT SOLUTIONS.

| Group | Start | End | Failure Event Count | | | | | | | | | | |
|---|---|---|---|---|---|---|---|---|---|---|---|---|---|
| | | | Topology 97953 | Topology 62711 | Topology 9843 | Topology 79140 | Topology 33716 | Topology 71823 | Topology 72851 | Topology 64169 | Topology 96962 | Baseline | *L3-follows-L1* |
| 0 | 0 | 0 | 8252 | 8250 | 8241 | 8232 | 8234 | 8229 | 8210 | 8200 | 8136 | 8429 | 8945 |
| 1 | 1 | $H$ | 3012 | 3007 | 3008 | 3016 | 3016 | 3019 | 3016 | 3019 | 3054 | 2760 | 2445 |
| 2 | $H+1$ | $2H$ | 796 | 801 | 807 | 805 | 803 | 805 | 820 | 816 | 843 | 842 | 682 |
| 3 | $2H+1$ | $3H$ | 25 | 27 | 29 | 32 | 32 | 32 | 39 | 46 | 49 | 42 | 13 |
| 4 | $3H+1$ | $4H$ | 2 | 2 | 2 | 2 | 2 | 2 | 2 | 6 | 5 | 11 | 2 |
| 5 | $4H+1$ | $5H$ | 2 | 2 | 2 | 2 | 2 | 2 | 2 | 2 | 2 | 5 | 2 |

| | | | | | | | | | | | |
|---|---|---|---|---|---|---|---|---|---|---|---|
| Vulnerability Metric | 0.0179 | 0.0179 | 0.0179 | 0.0180 | 0.0180 | 0.0180 | 0.0183 | 0.0184 | 0.0224 | 0.0201 | 0.0067 |
| Implementation Cost (normalized) | 0.4508 | 0.4218 | 0.4140 | 0.4125 | 0.4093 | 0.4032 | 0.4010 | 0.3923 | 0.3873 | 0.4138 | 1.0000 |

## V. Discussion and Future Work

### A. Choice of the Metaheuristic Method

Any multi-objective optimizer operating on binary decision vectors can be used at the outer level of ASCEND. To make the evaluation credible, multiple multi-objective optimizers must be compared because no single algorithm is uniformly best across problem classes [28]. We therefore compared three well-known algorithms from complementary paradigms: (i) MOPSO, a swarm-based method that uses Pareto dominance and an external archive and is widely adopted for multi-objective particle swarm optimization [9]; (ii) Non-dominated Sorting Genetic Algorithm II (NSGA-II), a canonical dominance-based evolutionary algorithm using elitist non-dominated sorting and crowding distance, widely used as a benchmark baseline [29]; and (iii) Multi-Objective Evolutionary Algorithm based on Decomposition (MOEA/D), a decomposition-based method that optimizes multiple scalar subproblems simultaneously, providing a fundamentally different mechanism for maintaining solution spread and convergence [30]. Collectively, these choices span swarm-based multi-objective optimizers and the two major Multi-Objective Evolutionary Algorithm (MOEA) search-strategy families—dominance-based and decomposition-based approaches [31].

We compared the outer-level metaheuristics by evaluating them on Market A with a sample size of 1000 over 50 iterations. All experimental conditions were held constant across runs; the only factor varied was the outer-level optimization algorithm (MOPSO, NSGA-II, or MOEA/D). The resulting Pareto fronts are shown in Fig. 12. Because MOPSO produced a Pareto front that dominated those obtained by the other methods (i.e., it yielded non-dominated solutions with no worse objective values and strictly better values in at least one objective), we selected MOPSO as the outer-level search method in ASCEND.

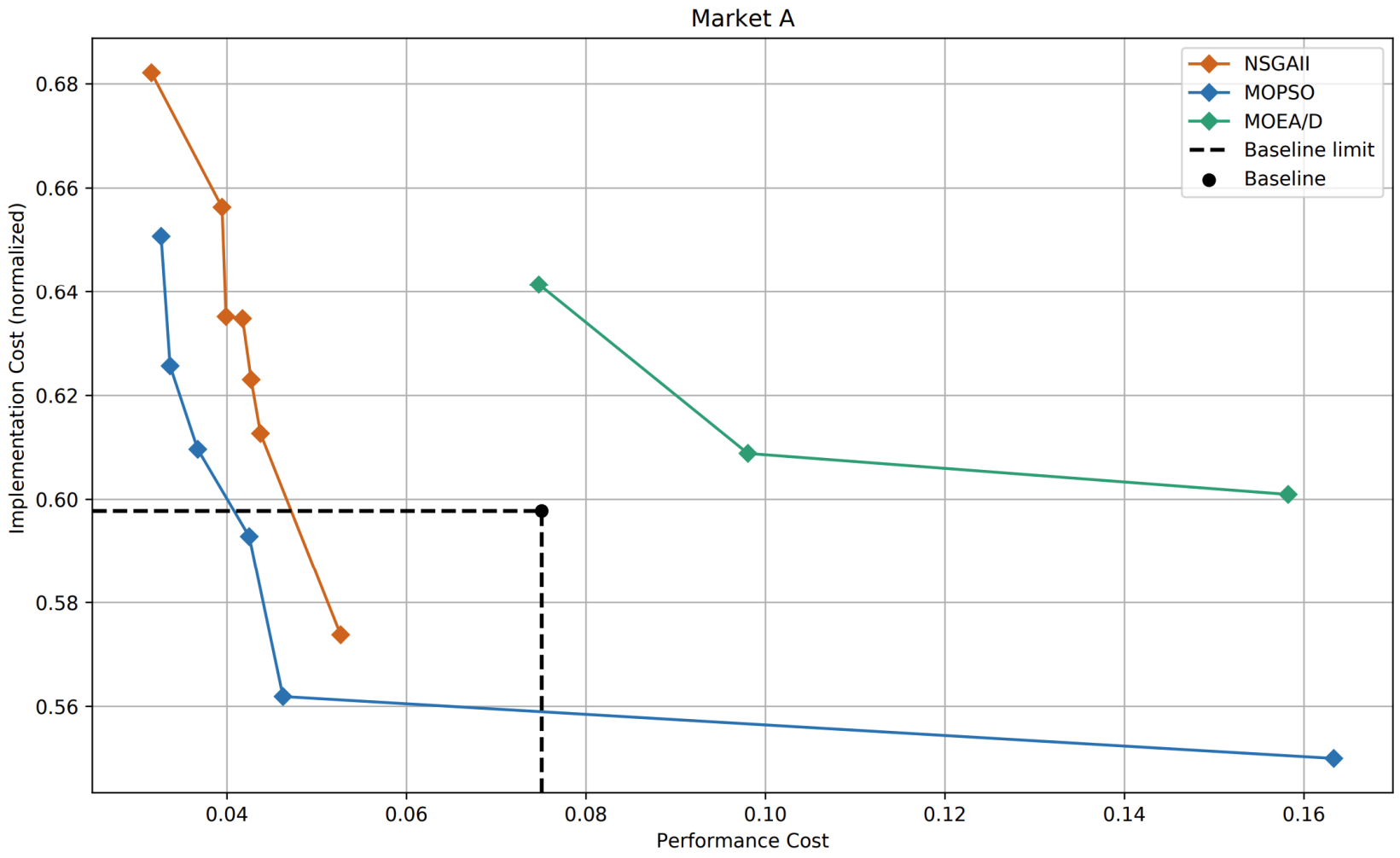

Fig. 12. Comparison of metaheuristic techniques for outer level search of ASCEND.

### B. Sensitivity to Optimization Budget

Metaheuristic optimization presents a direct tradeoff between solution quality and computational budget. In ASCEND, this tradeoff is particularly salient because each outer-loop MOPSO candidate requires solving an inner embedding problem, making long runs expensive in wall-clock time. For practical deployments we therefore cap the optimization at a fixed number of iterations (Table I), which ranges from 50 to 60 iterations in our evaluation and results in end-to-end run times of 2 to 7 days depending on market size.

To assess how the Pareto front evolves under a fixed compute budget—and to inform early-stopping decisions—we visualize the algorithm's intermediate behavior by plotting snapshots of the current non-dominated set at regular intervals. Concretely, for each market we run MOPSO for the configured maximum number of iterations, and every ten iterations (*i.e.*, at iterations $t \in \{10,20,30,\ldots\}$) we collect all candidate designs evaluated up to $t$ iterations and compute the non-dominated set with respect to the two objectives (performance cost and implementation cost). We then plot these non-dominated solutions in the 2D objective space, overlaying successive snapshots to visualize how the Pareto front approximation progresses over time (Fig. 13). To make the region of practical interest visible, we follow the same presentation choice used elsewhere in the paper and display only solutions whose costs fall below specified thresholds, focusing the visualization on the low-cost region where candidate designs would realistically be selected. Across markets, these "every-10-iterations" snapshots show that the plotted fronts increasingly overlap in later iterations, indicating that many improvements are achieved early and that additional iterations often yield diminishing returns within the low-cost region shown.

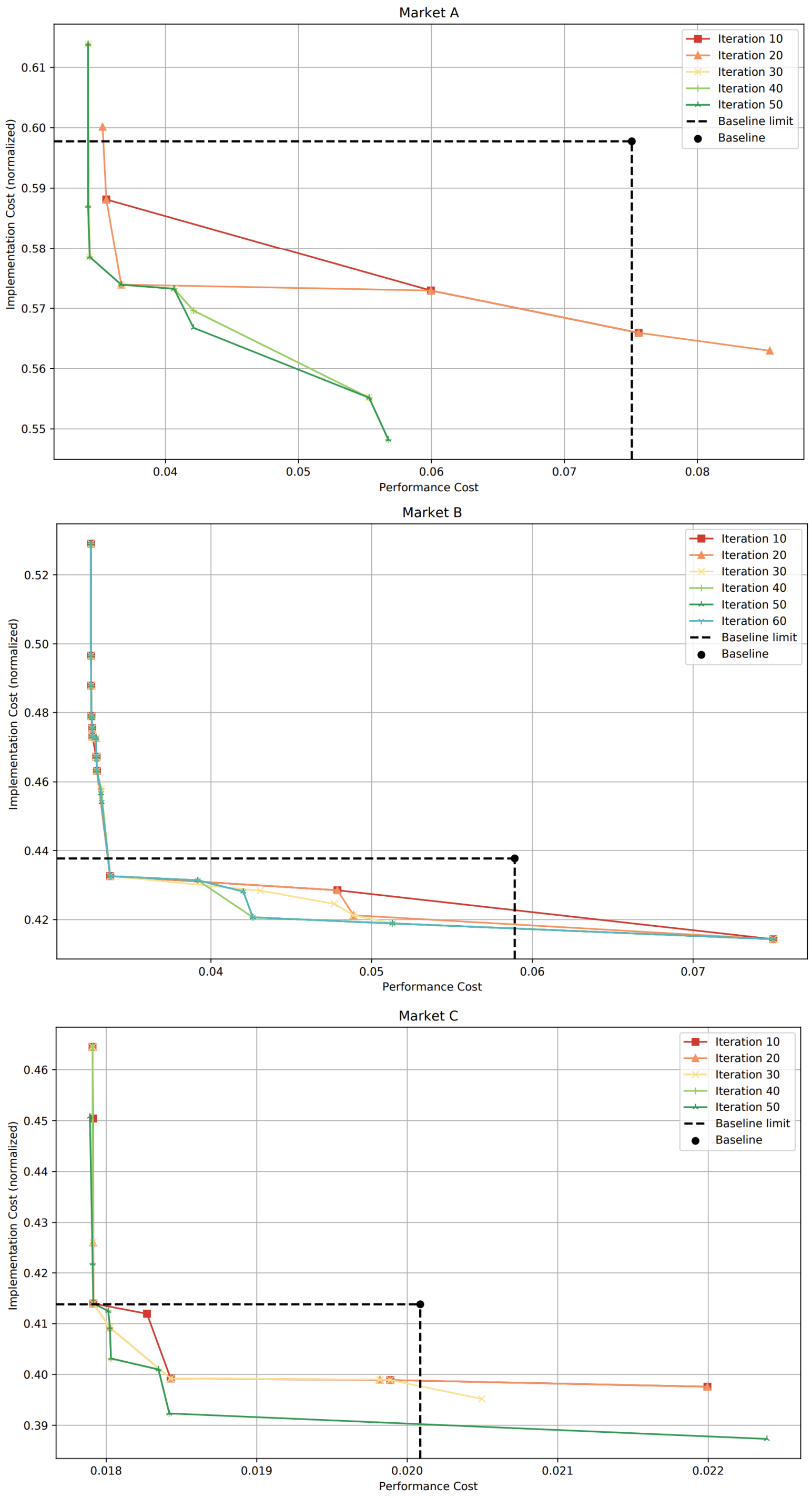

Fig. 13. Pareto front evolution. At each snapshot (every 10 iterations), we compute the non-dominated set among all designs evaluated up to that iteration and plot only solutions within the displayed cost thresholds to emphasize the low-cost region. Top to bottom: Market A, Market B, Market C.

The snapshot methodology above provides a practical early-stopping signal: when successive non-dominated snapshots (*e.g.*, $t$ vs. $t + 10$) exhibit negligible movement in the low-cost region, the run can be terminated to save compute while still returning a high-quality set of tradeoff designs. More formal adaptive stopping rules—such as stopping after a window of iterations with no meaningful improvement in a scalar front-quality indicator (*e.g.*, hypervolume stagnation [32])—are a natural extension; however, in this study we use fixed iteration budgets to ensure predictable run time across markets. It is also possible to save computational budget by running ASCEND with a smaller sample size, which we present in Appendix E.

### C. Penalty in Vulnerability Metric

In Section III-F1a, we introduced a vulnerability metric with a penalty that grows exponentially with the number of isolated homes for each outage. This metric is designed to compare candidate designs consistently and to align optimization with large-outage aversion; it is not intended as a calibrated predictor of absolute outage frequency.

The choice of the exponential base reflects how undesirable a single large outage is relative to multiple smaller outages. Using a higher base results in disproportionately larger penalties for large outages compared to smaller outages. For the experiments presented in this study, we selected a base value of 1.1 to provide increasing sensitivity to higher-impact outage scenarios. Under this parameterization, doubling outage magnitude produces a greater-than-linear increase in penalty. Under this penalty function, an outage affecting approximately 73 k homes results in a penalty of about 1.00, whereas an outage affecting 146 k homes yields a penalty of approximately 3.02. In other words, doubling the number of isolated homes increases the penalty by a factor of 3.02 (see Fig. 5). By contrast, if the penalty were defined as linearly proportional to the number of isolated homes ($Penalty\left(N_{DC}(event)\right) = N_{DC}(event)$), there would be no distinction between two outages affecting 10 k homes each and a single outage affecting 20 k homes in total.

More generally, future research could explore adaptive penalty functions that evolve based on historical outage data, customer impact metrics, or regulatory requirements. Machine-learning techniques could be used to learn penalty parameters directly from operational data, allowing the optimization process to better align with changing business objectives and service-level agreements.

### D. Alternative Failure Modeling

While this work focuses on fiber cuts, additional dimensions of failure modeling could be explored, such as site-level failures, device-level failures, and hybrid combinations thereof. Scenario-based or stress-testing approaches could also be incorporated to systematically evaluate network behavior under extreme but plausible conditions. In addition, failure models could account for non-uniform failure likelihoods, as certain fibers may be more prone to faults than others due to factors such as greater exposure to storms and hurricanes.

Furthermore, integrating time-dependent failure and repair processes would enable the framework to capture outage durations and restoration dynamics, including longer repair times for remote or difficult-to-access fibers. Such temporal modeling would provide a more comprehensive assessment of network resiliency by accounting not only for failure impact but also for recovery performance.

### E. Computational Complexity of ASCEND

From a systems perspective, the nested optimization structure can be computationally expensive at large scale because each outer-loop candidate requires solving an inner embedding problem. In addition to basic remedies, such as parallelization and archiving the cost values for repeating solutions, we employed other strategies to mitigate this issue as much as possible.

Based on our problem formulation, given $N$ possible core sites, MOPSO must optimize a binary vector of length $\frac{N(N-1)}{2}$, which theoretically results in $2^{\left(\frac{N(N-1)}{2}\right)}$ possible vectors. Of course, not all of these vectors correspond to distinct network topologies, due to overlaps among core circuits. To address this, we apply pruning to remove redundant core circuits and partially mitigate the high complexity (see Section III-D).

In addition, many sites offer little practical benefit as core sites. For example, some have low connectivity in the Layer-1 graph, while others are physically distant from neighboring sites, leading to high Layer-1 latency. Therefore, to reduce the dimensionality of the problem and avoid exploring unproductive regions of the search space, it is preferable to manually preselect a set of candidate sites that can serve as possible core sites. With sufficient computational resources, this candidate set can be expanded to explore a larger number of potential core sites. On the other hand, when there is sufficient confidence in the optimal selection of core sites based on prior knowledge, the choice of core sites can be fixed, and the optimization can be limited to the core circuits only. This approach further reduces the vector length by enabling circuit pruning to be applied only once before running the optimization pipeline (see Section III-D).

### F. Future Work

Future work could explore hierarchical or decomposed optimization strategies for the outer level, where the problem is solved at multiple abstraction levels—for example, regional subgraphs optimized independently and then coordinated at a higher layer. Hybrid approaches combining metaheuristics with exact or greedy methods may further improve scalability.

## VI. Conclusion

We presented ASCEND, an industry-scale framework for automated Layer-3 network design over an existing Layer-1 physical infrastructure under realistic correlated-failure assumptions. ASCEND frames the problem as a two-level multi-objective optimization task: an outer MOPSO search explores candidate core designs (core-site and core-circuit selections) while balancing resiliency and implementation cost, and an inner MCF formulation constructs a physically realizable end-to-end Layer-3 network realization by mapping both core and access connectivity onto Layer-1 paths that prioritize link/node/SRLG disjointness and low latency.

A key contribution is the performance cost model, which evaluates probabilistic failure scenarios—including SRLG-correlated events—using an exponential outage penalty together with explicit infeasibility penalties that enforce practical engineering constraints. Across multiple real-world market scenarios, ASCEND consistently identifies Pareto-front designs that outperform a manually engineered baseline in both resiliency and implementation cost, while matching or closely approaching the same higher-impact failure profile as *L3-follows-L1*. In particular, ASCEND can eliminate severe correlated-failure outcomes by selecting alternative core sites and circuit adjacencies that improve physical diversity. Future work includes extending the failure model to distribution-aware and time-dependent events, learning penalty parameters from operational data, and improving scalability.

## Appendix

### A. Search Space Magnitude

We analyze the size of the space of design vectors introduced in Section III-C. Because different vectors may yield identical realizations after pruning and embedding (Fig. 3), the difficulty of the outer-level search, and the cost of exhaustive enumeration, are governed by the size of this vector space.

Since each potential edge may be present or absent, a naïve upper bound is the number of possible undirected simple graphs on N labeled nodes. Let's call this number $G_N$:

$$G_N = 2^{\binom{N}{2}} = 2^{\frac{N(N-1)}{2}}$$

However, this estimate includes configurations that violate problem-specific constraints. In particular, our formulation requires that a predefined subset of sites must always function as mandatory core sites. Any graph in which these required sites are not part of the core (e.g., isolated or disconnected from the operational topology) is inherently invalid and should be excluded from consideration. Note that since the ASBR sites are always required core sites, connectivity of this subset also ensures the connectivity to the backbone.

To account for this constraint, we restrict the count to graph realizations in which all required core sites are actively present and participate in the core network. More specifically, we must count the number of graph configurations in which these mandatory core sites are guaranteed to be mutually reachable within the topology. We now quantify the size of the design space for core network configurations under this connectivity constraint. Let $R \subseteq V_c$ be a predefined subset of size M corresponding to required core sites. Connectivity between required core sites may be established either directly or via intermediate (non-required core) nodes.

We aim to count the number of simple undirected graphs on N labeled nodes that satisfy this constraint. Let this number be called $Q_{N,M}$, and let $C_N$ denote the number of connected labeled graphs on N nodes. The values of $C_N$ can be computed exactly using the recurrence:

$$C_N = G_N - \frac{1}{N}\sum_{k=1}^{N-1} k\binom{N}{k} C_k\, G_{N-k}, C_1 = 1.$$

In any qualifying graph, the required sites $R$ belong to one connected component of size $k$, where $M \leq k \leq N$. The component contains $k - M$ sites selected from the $N - M$ optional sites, giving $\binom{N-M}{k-M}$ choices. Its induced subgraph must be connected, giving $C_k$ possibilities, while the remaining $N - k$ sites form an arbitrary graph, giving $G_{N-k}$ possibilities. No edge can connect these two parts, since otherwise the component containing $R$ would be larger. Summing over all component sizes gives:

$$Q_{N,M} = \sum_{k=M}^{N} \binom{N-M}{k-M} C_k\, 2^{\binom{N-k}{2}}$$

This expression exactly counts graphs satisfying the required-core connectivity condition, but remains an upper bound on the fully feasible space because degree constraints (Section III-F1b) are not enforced and optional sites may form disconnected core components.

Table V reports $G_N$, $Q_{N,M}$, and the unique topologies evaluated by ASCEND. The evaluated count is at most the product of the particle count and iteration count because MOPSO may revisit a vector. The required-core constraint removes only 0.2–5% of configurations in these markets, leaving the search space on the same order as $G_N$. Exhaustive enumeration of the unpruned spaces would take years based on the observed evaluation times. With fixed core sites, pruning reduces the space to $2^{len/p}$(Section III-D), but the space still grows exponentially with the number of candidate sites.

TABLE V
SEARCH MAGNITUDE AND RUN DETAILS.

| Market | # of core site candidates (N) | # of required core sites (M) | # all possible core-site/circuit combinations ($G_N$) | # combinations that satisfy required core condition ($Q_{N,M}$) | # unique tested topologies in ASCEND | ASCEND wall clock |
|---|---|---|---|---|---|---|
| **Market A** | 13 | 10 | $\approx 3.02 \times 10^{23}$ | $\approx 3.01 \times 10^{23}$ | 100000 | 2D, 1H, 32M |
| **Market B** | 8 | 6 | 268435456 | 255334512 | 478233 | 6D, 6H, 41M |
| **Market C** | 9 | 9 | 68719476736 | 66296291072 | 99998 | 7D, 2H, 55M |

*B. All Topologies*

Fig. 14 shows all topologies evaluated by ASCEND, including those outside the practical low-cost region reported in Section IV; most of these additional topologies incur infeasibility penalties and are not deployable.

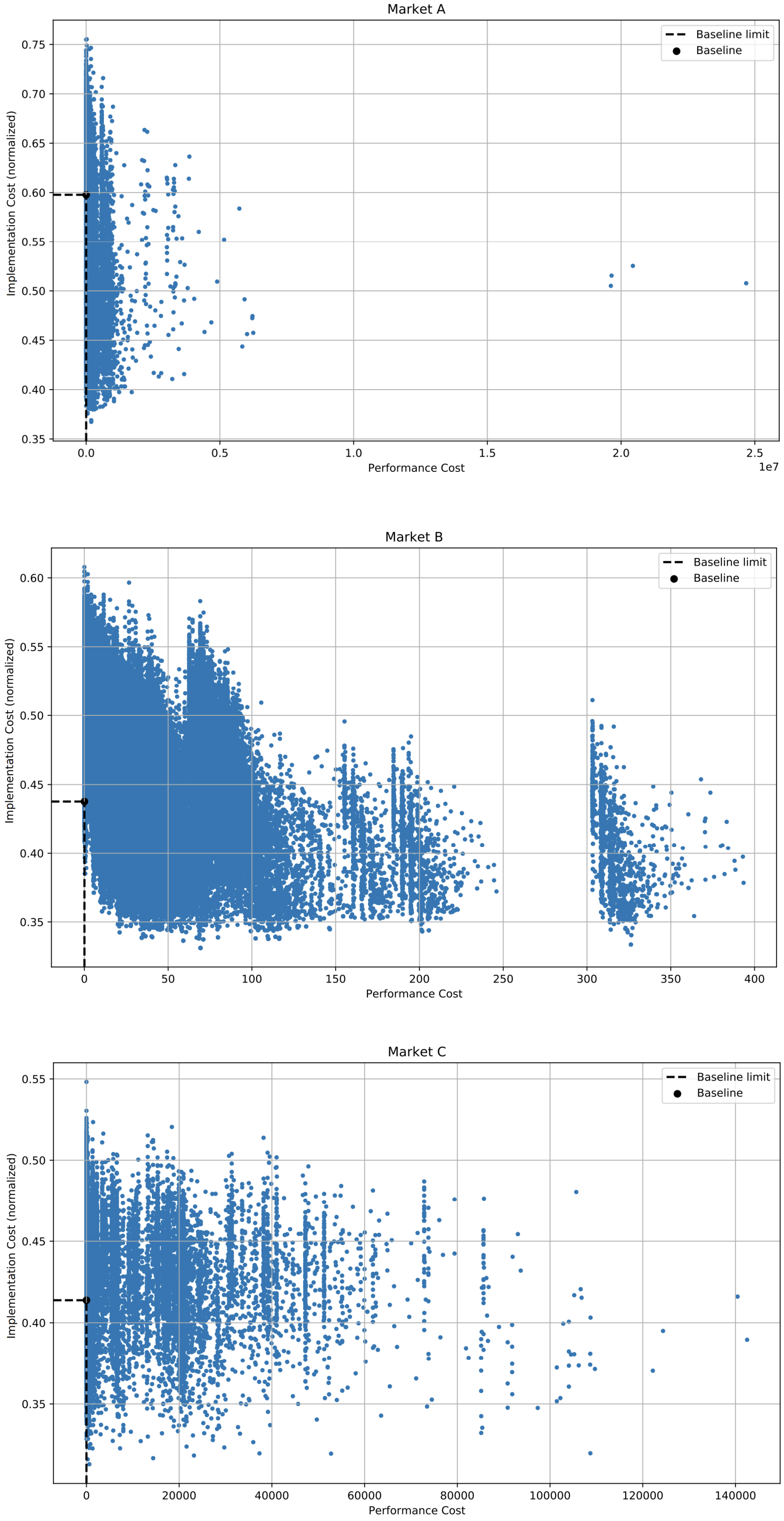

Fig. 14. All discovered topologies. Top to bottom: Market A, Market B, Market C.

### C. Seed Sensitivity

To evaluate MOPSO's stochastic variability, we ran ASCEND on Market A three times using independent seeds, 1000 particles, and 50 iterations. The runs produce broadly similar trade-off regions despite local differences in their fronts (Fig. 15). The main text reports one seed fixed in advance rather than selected post hoc.

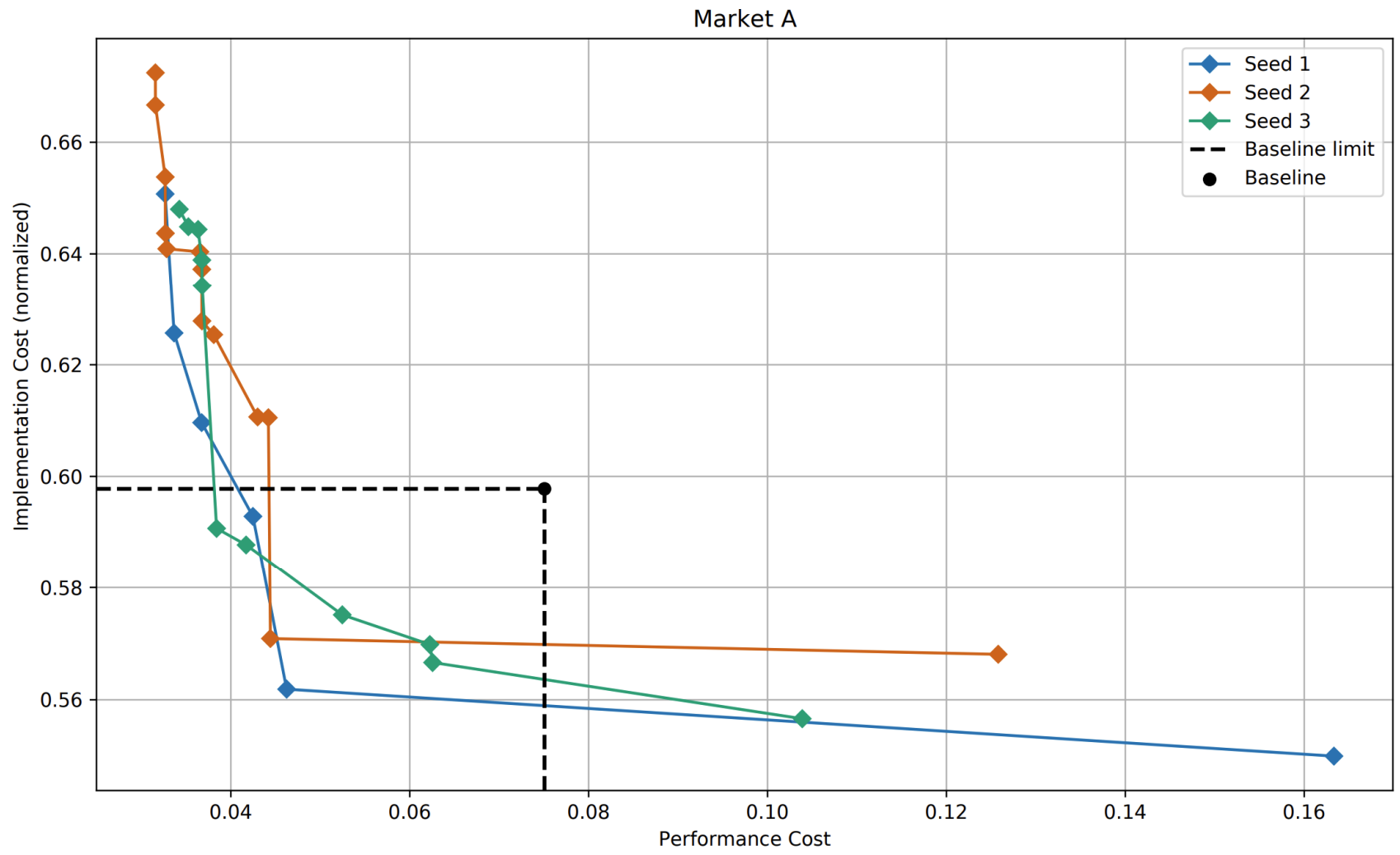

Fig. 15. Pareto front of different seeds in MOPSO for Market A.

### D. Baseline Topology

The baseline topologies were produced using a legacy semi-automated network-design process that combines computational analysis with engineering review. The process consists of three general stages: selection of candidate core sites, construction and refinement of Layer-3 connectivity, and evaluation of the resulting topology under the resiliency and implementation-cost criteria considered in this work. Implementation-specific details of the baseline process are omitted to protect proprietary and business-sensitive information, while preserving the methodological information necessary to support the comparisons presented in this work.

Candidate core configurations are first evaluated using resiliency analysis, after which promising configurations are selected for further evaluation. Layer-3 connectivity is then constructed and iteratively refined to remove unnecessary connectivity while maintaining predefined engineering and resiliency requirements. The resulting designs are evaluated under the same general failure-analysis framework used for comparison with ASCEND, followed by engineering review and, where appropriate, additional design refinements.

This baseline approach does not exhaustively explore all possible combinations of core-site and core-circuit decisions. Instead, it progressively narrows the design space using engineering criteria and intermediate analysis. Consequently, the quality of the resulting topology can depend on the sequence of design decisions and engineering review. ASCEND addresses this limitation by systematically exploring the joint core-site and core-circuit design space through the multi-objective optimization framework described in Section III.

### E. Effect of Sample Size

We evaluated ASCEND on Market A using sample sizes (number of MOPSO particles) of 200, 1000, 2000, and 5000 while holding the iteration count at 50. Table VI reports the corresponding run times. As shown in Fig. 16, increasing the sample size shifts the front toward lower implementation cost and improves consistency within the practical region. A sample size of 200 produces relatively few solutions below the baseline cost limit; 1000 provides a substantial improvement, while 2000 further tightens the solution cluster. Increasing the sample size to 5000 provides no clear improvement over 2000. A sample size of 1000–2000 therefore offers a practical balance between solution quality and run time.

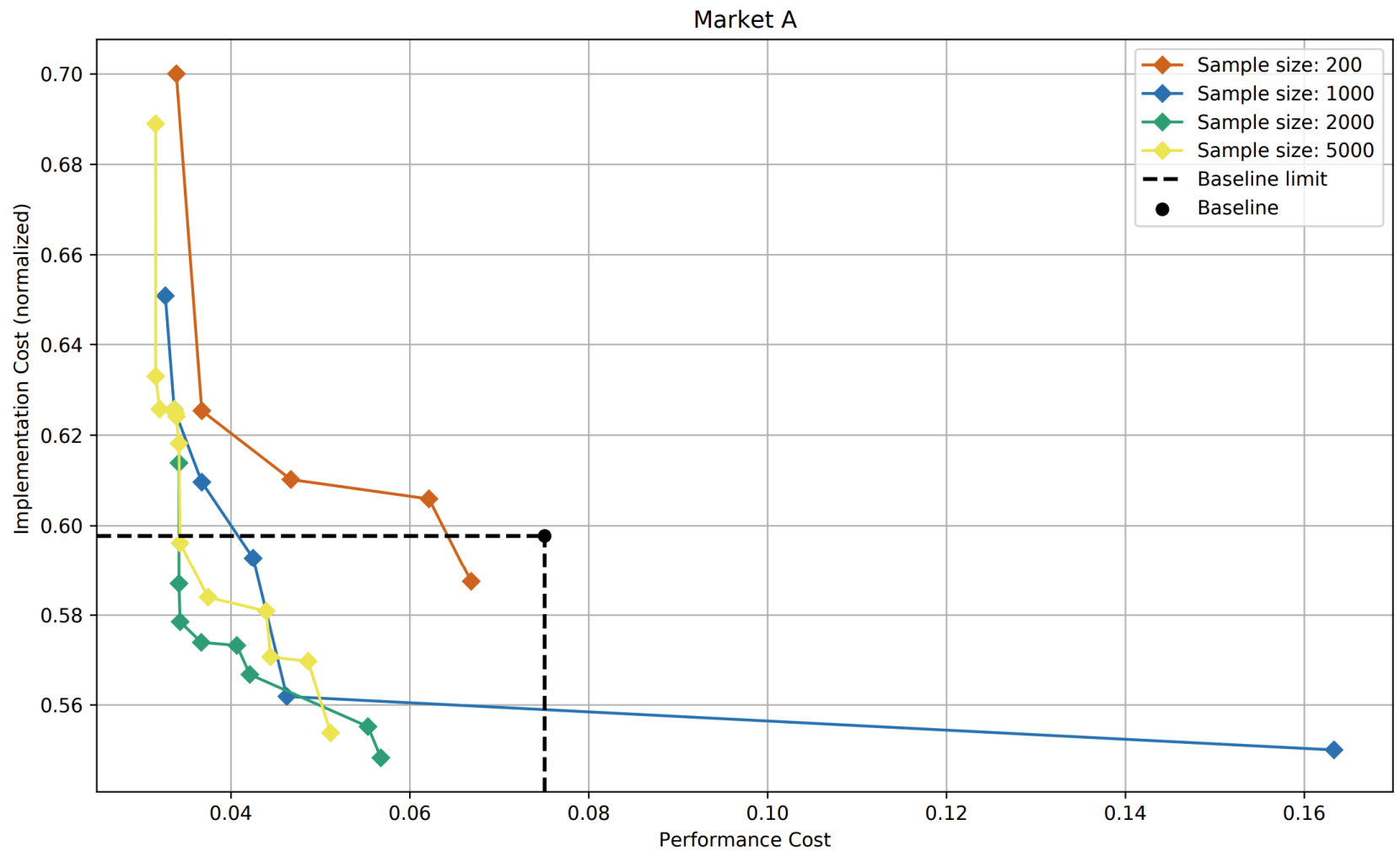

Fig. 16. Pareto front of different sample sizes in MOPSO for Market A.

TABLE VI
WALL CLOCK OF ASCEND WITH DIFFERENT MOPSO SAMPLE SIZES.

| **Sample size** | **Wall clock** |
|---|---|
| 200 | 5H, 44M |
| 1000 | 1D, 4H, 7M |
| 2000 | 2D, 1H, 32M |
| 5000 | 5D, 22H, 59M |